\documentclass[10pt]{article}
\pdfoutput=1

\usepackage{bm,wrapfig,float,array,mathtools}
\usepackage{amsfonts}
\usepackage{amssymb}
\usepackage{cite}
\usepackage{amsmath}
\usepackage{color}
\usepackage{graphicx}
\usepackage{hyperref}
\usepackage{float}
\usepackage[T1]{fontenc}
\usepackage[utf8]{inputenc}
\usepackage{alphabeta}
\usepackage{fancyvrb}
\usepackage[dvipsnames]{xcolor}
\usepackage{bold-extra}
\usepackage{lmodern}
\usepackage{cleveref}
\usepackage[normalem]{ulem}
\usepackage{tikz-cd}
\usepackage{tensor}
\usepackage{multirow}

\usepackage[all]{xy}

\hypersetup{colorlinks=true, citecolor= NavyBlue,
  linkcolor= Maroon, urlcolor=NavyBlue}

\usepackage{colortbl}

\graphicspath{ {figures/} }

\usepackage{tikz}
\usetikzlibrary{arrows}
\usetikzlibrary{shapes.geometric,calc,arrows, positioning,shapes.misc,decorations.markings}
\tikzset{
  big arrow/.style={
    decoration={markings,mark=at position 1 with {\arrow[scale=2,#1]{>}}},
    postaction={decorate},
    shorten >=0.4pt},
  big arrow/.default=black}
\usetikzlibrary{external}
\usetikzlibrary{positioning}
\usetikzlibrary{calc}
\tikzset{gauge-node/.style={shape=circle, draw, minimum width=.6cm}}
\tikzset{global-node/.style={shape=rectangle, draw, minimum width=.6cm}}

\tikzstyle{brane}=[draw]
\tikzset{D7/.style={circle, draw=black, inner sep=0pt, fill=white, minimum size=3mm}}
\tikzset{hasse/.style={circle, fill,inner sep=2pt}}
\tikzset{flavor/.style={regular polygon,regular polygon sides=4,inner sep=2.5pt, draw}}
\tikzset{gauge/.style={circle, draw,inner sep=2.5pt}}
\tikzset{gaugeb/.style={circle, draw,fill=black,inner sep=2.5pt}}
\tikzset{gaugecyan/.style={circle, draw,fill=cyan,inner sep=2.5pt}}
\tikzset{gaugegreen/.style={circle, draw,fill=green,inner sep=2.5pt}}
\tikzset{gaugeblue/.style={circle, draw,fill=blue,inner sep=2.5pt}}
\tikzset{gaugeorange/.style={circle, draw,fill=orange,inner sep=2.5pt}}
\tikzset{fd/.style={circle, draw,fill= ForestGreen ,inner sep=2.5pt}}
\tikzset{bd/.style={circle, draw=black, inner sep=0pt, fill=black, minimum size=2mm}}
\tikzset{wd/.style={circle, draw=black, inner sep=0pt, fill=white, minimum size=2mm}}
\tikzset{Dynkin/.style={circle, draw=black, inner sep=0pt, fill=white, minimum size=2mm}}
\tikzstyle{ligne}=[draw, thick] 
\tikzstyle{lignew}=[draw=purple, thick] 
\tikzstyle{lignet}=[draw=orange] 
\tikzset{doublearrow/.style={ draw=black!75, color=black!75, thick, double distance=3pt, }}

\pgfdeclarelayer{edgelayer}
\pgfdeclarelayer{nodelayer}
\pgfsetlayers{edgelayer,nodelayer,main} 
\tikzstyle{none}=[inner sep=0pt] 

\tikzstyle{NodeCross}=[draw, shape=circle, cross out, inner sep=0pt, minimum size=6pt,line width=0.25mm]
\tikzstyle{Circle}=[draw, shape=circle, black, inner sep=0pt, minimum size=6pt]
\tikzstyle{rtriangle}=[fill=black, regular polygon, regular polygon sides=3, rotate=90, inner sep=0pt, minimum size=8pt]
\tikzstyle{ltriangle}=[fill=black, regular polygon, regular polygon sides=3, rotate=270, inner sep=0pt, minimum size=8pt]
\tikzstyle{rtriangleblue}=[fill={rgb,255: red,17; green,160; blue,255}, regular polygon, regular polygon sides=3, rotate=90, inner sep=0pt, minimum size=8pt]
\tikzstyle{ltriangleblue}=[fill={rgb,255: red,17; green,160; blue,255}, regular polygon, regular polygon sides=3, rotate=270, inner sep=0pt, minimum size=8pt]
\tikzstyle{rtrianglegreen}=[fill={rgb,255: red,69; green,255; blue,28}, regular polygon, regular polygon sides=3, rotate=90, inner sep=0pt, minimum size=8pt]
\tikzstyle{ltrianglegreen}=[fill={rgb,255: red,69; green,255; blue,28}, regular polygon, regular polygon sides=3, rotate=270, inner sep=0pt, minimum size=8pt]
\tikzstyle{Uprtriangle}=[fill=black, regular polygon, regular polygon sides=3, rotate=0, inner sep=0pt, minimum size=8pt]
\tikzstyle{Downltriangle}=[fill=black, regular polygon, regular polygon sides=3, rotate=180, inner sep=0pt, minimum size=8pt]
\tikzstyle{rtriangleAmber}=[fill={rgb,255: red, 191; green, 144; blue, 63}, regular polygon, regular polygon sides=3, rotate=90, inner sep=0pt, minimum size=8pt]
\tikzstyle{UprtriangleViolett}=[fill={rgb,255: red,255; green,0; blue,0}, regular polygon, regular polygon sides=3, rotate=0, inner sep=0pt, minimum size=8pt]

\tikzstyle{Downltriangle}=[fill=black, regular polygon, regular polygon sides=3, rotate=180, inner sep=0pt, minimum size=8pt]
\tikzstyle{UpRighttriangle}=[fill=black, regular polygon, regular polygon sides=3, rotate=45, inner sep=0pt, minimum size=8pt]
\tikzstyle{UpLefttriangle}=[fill=black, regular polygon, regular polygon sides=3, rotate=315, inner sep=0pt, minimum size=8pt]
\tikzstyle{DownRighttriangle}=[fill=black, regular polygon, regular polygon sides=3, rotate=135, inner sep=0pt, minimum size=8pt]
\tikzstyle{DownLighttriangle}=[fill=black, regular polygon, regular polygon sides=3, rotate=225, inner sep=0pt, minimum size=8pt]

\tikzstyle{Star}=[draw, shape=star, fill=black, star points=8, inner sep=0pt, minimum size=8pt]

\tikzstyle{DashedLine}=[-, densely dashed, line width=0.25mm]
\tikzstyle{DashedLineBrown}=[-, densely dashed, line width=0.25mm, draw={rgb,255: red,155; green,103; blue,51}]
\tikzstyle{DashedLineFall}=[-, densely dashed, line width=0.25mm, draw={rgb,255: red,195; green,0; blue,0}]
\tikzstyle{DashedLineViolett}=[-, densely dashed, line width=0.25mm, draw={rgb,255: red,139; green,41; blue,148}]
\tikzstyle{DottedLine}=[-, dotted, line width=0.25mm]
\tikzstyle{BlueLine}=[-, fill=none, draw={rgb,255: red,17; green,160; blue,255}, line width=0.25mm]
\tikzstyle{GreenLine}=[-, fill=none, draw={rgb,255: red,69; green,255; blue,28}, line width=0.25mm]
\tikzstyle{RedLine}=[-, draw={rgb,255: red,191; green,0; blue,0}, fill=none, line width=0.25mm]
\tikzstyle{DashedLineRed}=[-, densely dashed, fill=none, draw={rgb,255: red,191; green,0; blue,0}, line width=0.25mm]
\tikzstyle{ThickLine}=[-, line width=0.25mm]
\tikzstyle{ViolettLine}=[-, draw={rgb,255: red,132; green,60; blue,191}, fill=none, line width=0.25mm]
\tikzstyle{ViolettDashedLine}=[-, densely dashed, draw={rgb,255: red,132; green,60; blue,191}, fill=none, line width=0.25mm]
\tikzstyle{AmberLine}=[-, draw={rgb,255: red,191; green,144; blue,63}, fill=none, line width=0.25mm]
\tikzstyle{DashedRedThick}=[-, densely dashed, fill=none, draw={rgb,255: red,191; green,0; blue,0}, line width=0.40mm]
\tikzstyle{DashedBlueThick}=[-, densely dashed, fill=none, black, line width=0.40mm]

\tikzstyle{blackdot}=[fill=black, draw=black, shape=circle]
\tikzstyle{pq-brane}=[-, draw=red, thick]
\tikzstyle{BOUNDARY}=[-, draw={rgb,255: red,255; green,184; blue,19},  thick]
\tikzstyle{EXCISION}=[-, draw={rgb,255: red,40; green,170; blue,116}, fill={rgb,255: red,40; green,170; blue,116}, 
tikzit draw={rgb,255: red,40; green,170; blue,116}, tikzit fill={rgb,255: red,40; green,170; blue,116}]

\newcommand{\bea}{\begin{eqnarray}}
\newcommand{\eea}{\end{eqnarray}}
\newcommand{\be}{\begin{equation}}
\newcommand{\ee}{\end{equation}}
\newcommand{\ba}{\begin{aligned}}
\newcommand{\ea}{\end{aligned}}
\newcommand{\bit}{\begin{itemize}}
\newcommand{\eit}{\end{itemize}}
\newcommand{\ben}{\begin{enumerate}}
\newcommand{\een}{\end{enumerate}}

\newcommand{\beq}{\begin{equation}}
\newcommand{\eeq}{\end{equation}}
\def\be {\begin{equation}}
\def\ee {\end{equation}}
\def\bs#1\es{\begin{split}#1\end{split}}
\def\bg#1\eg{\begin{gathered}#1\end{gathered}}
\def\bea{\begin{eqnarray}}
\def\eea{\end{eqnarray}}

\newcommand{\SNF}{\text{SNF}}
\newcommand{\diag}{\text{diag}}
\newcommand{\Bock}{\text{Bock}}

\usepackage{soul}

\usepackage{customprelude}

\begin{document}

% format
\baselineskip=18pt  % a la harvmac
\numberwithin{equation}{section}  % make eq labels (sec.num)
\allowdisplaybreaks  % allow page breaks in displayed eqs

%%%%%%%%%%%%%%%%%%%%%%%%%%%%%%%%%%%%%%%%%%%
%%%        TITLE BEGINS HERE
%%%%%%%%%%%%%%%%%%%%%%%%%%%%%%%%%%%%%%%%%%%

%% ========== title (note version) begins here ==========

%\vspace*{-1cm}
%\begin{center}
% {\LARGE }
%\end{center}
%\vspace*{-.5cm}

% ========== title (note version) ends here ==========

% ========== title (paper version, a la harvmac) begins here ==========

\thispagestyle{empty}

% title, authors, affiliation

\vspace*{0.8cm} 
\begin{center}
{{\huge 
2-Group Symmetries and M-Theory
}}

 \vspace*{1.5cm}
Michele Del Zotto\,$^{1}$, 
\ I\~naki Garc\'ia Etxebarria\,$^2$,
\   Sakura Sch\"afer-Nameki\,$^{3}$\\

\bigskip
\smallskip

{\it $^1$ Mathematics Institute, Uppsala University, \\
Box 480, SE-75106 Uppsala, Sweden}\\

{\it $^1$ Department of Physics and Astronomy, Uppsala University, \\
Box 516, SE-75120 Uppsala, Sweden}

\smallskip

{\it $^2$ Department of Mathematical Sciences,\\
Durham University, Durham, DH1 3LE, United Kingdom}

\smallskip

{\it  $^3$ Mathematical Institute, University of Oxford, \\
Andrew-Wiles Building,  Woodstock Road, Oxford, OX2 6GG, UK}\\
\smallskip

\end{center}

\vspace*{0.8cm}

\noindent
Quantum Field Theories engineered in M-theory can have
2-group symmetries, mixing 0-form and 1-form symmetry backgrounds in
non-trivial ways. In this paper we develop methods for determining the
2-group structure  from the boundary geometry of
the M-theory background. We illustrate these methods in the case of 5d
theories arising from M-theory on ordinary and generalised toric
Calabi-Yau cones, including cases in which the resulting theory is
non-Lagrangian. Our results confirm and elucidate previous results on 2-groups from geometric engineering.

\newpage
%%%%%%%%%%%%%%%%%%%%%%%%%%%%%%%%%%%%%%%%%%%
%%%           TITLE ENDS HERE
%%%%%%%%%%%%%%%%%%%%%%%%%%%%%%%%%%%%%%%%%%%

\tableofcontents
%\printindex

%%%%%%%%%%%%%%%%%%%%%%%%%%%%%%%%%%%%%%%%%%%
%%%        MAIN TEXT BEGINS HERE
%%%%%%%%%%%%%%%%%%%%%%%%%%%%%%%%%%%%%%%%%%%
\section{Introduction}

A modern understanding of symmetries characterizes them as a subsector of
topological\footnote{\, Meaning that their dependence on their support is topological (which ensures their conservation) up to collisions with charged operators.}
operators with support of various codimensions \cite{Kapustin:2013uxa,Gaiotto:2014kfa}.  The codimension in
spacetime determines the dimensions of the charged (not necessarily
topological) extended operators in the theory. If a quantum field
theory (QFT) has a symmetry described by topological operators of different codimension, these
can have non-trivial fusion rules, which mix defects of different
codimension. In the most general case, the resulting fusions give
rise to categorical symmetries
\cite{Bhardwaj:2017xup,Tachikawa:2017gyf,Thorngren:2019iar,Komargodski:2020mxz,Kaidi:2021xfk,Choi:2021kmx}. Sometimes,
categorical symmetries organize in so-called higher-groups
\cite{Kapustin:2013uxa,Sharpe:2015mja, Tachikawa:2017gyf,
  Cordova:2018cvg,Benini:2018reh,Cordova:2020tij}. These are interesting mathematical
structures that are ubiquitous in the landscape of quantum field
theories:\footnote{$\,$ We refer our readers to the important
  foundational papers
  \cite{Baez:2005sn,Sati:2008eg,Sati:2009ic,Fiorenza:2010mh,Fiorenza:2012tb}
  where the crucial interplay among the Green-Schwarz mechanism and
  2-groups (and string 2-Lie algebras) was originally derived.}
2-groups have been found in the most basic textbook examples of 4d
gauge theories, such as Quantum Electro-Dynamics (QED) and gauge
theories with matter \cite{Cordova:2018cvg, Lee:2021crt,
  Apruzzi:2021mlh}, as well as in some of the most exotic systems,
such as superconformal field theories (SCFTs) in 5d
\cite{Apruzzi:2021vcu} and 6d \cite{Apruzzi:2021mlh}, and little
string theories \cite{Cordova:2020tij,DelZotto:2020sop}.

Since the generic supersymmetric QFT does not admit a conventional
Lagrangian description, it is paramount to develop tools in string
theory to detect and study the features of such generalized symmetries
in the context of various geometric engineering  scenarios
\cite{Acharya:2001hq,DelZotto:2015isa,GarciaEtxebarria:2019caf, Morrison:2020ool, Albertini:2020mdx,
  Apruzzi:2020zot, Braun:2021sex, Closset:2020scj, DelZotto:2020sop,
  Bhardwaj:2020phs, Closset:2020afy, Apruzzi:2021vcu, Apruzzi:2021phx,Hosseini:2021ged,
  Apruzzi:2021mlh, Apruzzi:2021nmk, Bhardwaj:2021mzl,Buican:2021xhs, Cvetic:2021maf,
  Cvetic:2021sxm, Cvetic:2021sjm, Closset:2021lwy, Tian:2021cif, Genolini:2022mpi,
  DelZotto:2022fnw, Cvetic:2022uuu, HubnerMorrisonSSNWang}.\footnote{\ Following the geometric engineering paradigm of exploiting geometries to inform SQFTs \cite{Katz:1996fh,Leung:1997tw}.}  Indeed,
2-groups structures have been computed in the literature from
geometries engineering SCFTs and LSTs in various dimensions
\cite{DelZotto:2020sop,Apruzzi:2021phx,Apruzzi:2021vcu,Bhardwaj:2021wif,DelZotto:2022fnw}. Although
so far the cases of interest to us have only been studied on a
resolved phase of the geometry, it seems natural to expect that
2-groups should be an intrinsic feature of these setups, encoded in
the geometry of the singularity itself and not only in specific properties of its resolutions,
rulings, or deformations. This fact, together with the idea that the charge lattice of extended
operators in stringy constructions can be identified with relative
homology, indicates that higher groups must be captured by properties
of the link (i.e. boundary) of the singularity. For higher-form
symmetries this expectation is confirmed by e.g. analysing the
non-commutativity of fluxes at the link
\cite{Freed:2006yc,Freed:2006ya,GarciaEtxebarria:2019caf,Morrison:2020ool,Albertini:2020mdx,Closset:2020scj,DelZotto:2020esg,Hosseini:2021ged},
the structure of the symmetry TFTs arising from a reduction on the boundary of
the compactification \cite{Apruzzi:2021nmk}, and holographic analysis
\cite{Witten:1998wy, Bergman:2020ifi, Apruzzi:2021phx}.

In this paper we show that this is also the case for 2-groups by
deriving the 2-group symmetry from a boundary perspective. Concretely,
for geometries where the zero-form symmetries that act faithfully are
manifestly realized in terms of non-compact singularities, the
latter give rise to singularities in the link geometry. Our main
result is a derivation of the 2-group structure 
from the geometry
of the singular link, by relating it to the
structure of line operators as described in
\cite{Apruzzi:2021vcu,Bhardwaj:2021wif,Lee:2021crt}. We stress that this result is true in general,
irrespective of the dimensionality of the singularity in question.  We
first develop this {dictionary} in general, and then apply the resulting formalism to the
case of M-theory compactifications on three dimensional Calabi-Yau
(CY) singularities, which are dual to $(p,q)$ five-brane webs. We also
consider generalized toric models, which are dual to webs with
multiple 5-branes ending on a single 7-brane. For many of these models
the 5d SCFTs are known to have 2-group symmetries
\cite{Apruzzi:2021vcu}, and we confirm (and in part extend) these
results, using this boundary perspective. Clearly a more general
analysis of 2-groups for all 5d SCFTs beyond the (generalized) toric
models can be carried out. It requires a framework where the flavor
symmetry is manifest, either in terms of gluing of surfaces
\cite{Bhardwaj:2020avz, Bhardwaj:2020ruf}, non-flat resolutions
\cite{Apruzzi:2019vpe, Apruzzi:2019opn, Apruzzi:2019enx,
  Apruzzi:2019kgb} or orbifolds \cite{Benini:2009gi,Acharya:2021jsp,Tian:2021cif,DelZotto:2022fnw}. For instance, the case of the $\mathbb{F}_0$ description of the $SU(2)_0$ Seiberg theory is not amenable to this approach, whereas in contrast the $\mathbb{F}_2$ one, which has the manifest non-abelian flavor symmetry realized geometrically, is. 

\medskip

Field theoretically we can characterize higher-groups by studying the background fields for global symmetries that are generated by topological operators. In the case of 2-groups the 1-form symmetry and 0-form symmetry of a QFT satisfy a non-trivial relation: the variation of the background of the 1-form symmetry does not vanish, but depends on the 0-form symmetry background. An alternative but equivalent description of 2-groups emerges by considering equivalence relations on line operators induced by local
operators \cite{Apruzzi:2021vcu, Lee:2021crt, Bhardwaj:2021wif}. In particular the 1-form symmetry group is the (Pontryagin dual group to the) group of lines modulo the relation induced by local line changing operators. In the presence of local
operators charged under flavor symmetries, this screening picture can be refined by taking into account the behaviour under the global symmetry of the line changing operator, and considering only those relations induced by operators in (proper) representations of the
flavor group. 

Field theoretically\footnote{\, In {the context of} 5d {SCFTs} this is however not purely a field-theoretic analysis, since {in practice} the charges of operators including non-perturbative states such as instanton particles are computed through geometric methods.} this interplay between 0- and 1-form symmetry can be detected by computing the charges of local operators under gauge and flavor symmetries. It is this second characterization of the 2-group structure in terms of lines that we reproduce from geometry in this paper. All the various field theoretical ingredients translate nicely in geometrical properties of the boundary of the singularity, and more specifically into the interplay of the local flavor structure arising from the geometry close to the singular locus and the geometry of the rest of the link.

A question that we will not address in this paper is the following:
given a 5d theory with a 2-group, one can gauge the 1-form symmetry
and obtain a theory with a 0-form symmetry, a 2-form symmetry, and a
mixed 't Hooft anomaly connecting them \cite{Tachikawa:2017gyf}. We
expect, based on previous experience in the context of 1-form
symmetries (see for instance
\cite{Witten:1998wy,GarciaEtxebarria:2019caf,Morrison:2020ool,Albertini:2020mdx,Closset:2020scj,DelZotto:2020esg,Hosseini:2021ged})
that both possibilities will be realised in string theory, with the
choice being determined by a choice of boundary conditions for fluxes
at infinity. Relatedly, we expect that compactification of M-theory on
the link of the cone geometry, along the lines of the analysis in
\cite{Apruzzi:2021nmk}, leads to a topological field theory  -- the Symmetry TFT -- in one dimension higher. A choice of gapped interface in this theory
encodes the polarization choice, between having a 2-group and having ordinary
symmetries with a mixed anomaly (this assumes that we can gauge the 1-form symmetry in the 2-group, or the 2-form symmetry in the mixed anomaly, respectively). In this paper we will from the
beginning make a choice between these two possibilities by assuming
that the non-compact M2-branes lead to genuine line operators in the
field theory, which leads to 1-form symmetries in the field theory and
therefore 2-group structures. It would certainly be interesting to
understand the general situation, but we leave this for the future.

\medskip

The structure of this paper is as follows: We begin with a brief recap
of 2-group symmetries, summarizing their salient features in section
\ref{sec:physics}.  We then derive our main result, the 2-group
symmetries in M-theory geometric engineering from the boundary of the
compactification space, in section \ref{sec:core}.  Section
\ref{sec:5dcase} applies this general approach to M-theory
compactifications on singular Calabi-Yau three-folds to 5d SCFTs, and
we show the equivalence to other approaches using five-brane webs and the original
intersection theory computations \cite{Apruzzi:2021vcu}. We provide a flurry of examples, which we
discuss using these complementary approaches in section
\ref{sec:Exes}.

%%%%%%%%%%%%%%%%%%%%%%%%%%%%%
\section{Generalized Symmetries from The Boundary}\label{sec:generalcase}

\subsection{A Recap of 2-Group Symmetries}\label{sec:physics}

In this section we give a brief review of 2-group symmetries following \cite{Apruzzi:2021vcu, Bhardwaj:2021wif,Lee:2021crt} in order to fix notation and conventions. The 2-groups we consider here are built out of discrete 1-form symmetries and continuous 0-form symmetries
-- for a more general analysis of 2-groups, including continuous 1-form symmetries, see \cite{DelZotto:2020sop,Apruzzi:2021phx,Apruzzi:2021vcu,Bhardwaj:2021wif,DelZotto:2022fnw}. Consider a theory $\mathcal T$ with a discrete 1-form symmetry $\Gamma^{(1)}$ and a
continuous 0-form symmetry $\mathcal{F}^{(0)} = F/C$,
 where
$F$ is a simply-connected Lie group, and $C$ a subgroup of its center.
We define the
  global form of the flavor symmetry (0-form) group $\mathcal{F}^{(0)}$ as the group acting faithfully on
  the spectrum of local operators, or equivalently as the most general
  structure group that we can choose for the background fields.

The theory $\mathcal T$ has a collection of genuine line operators $\sL$. 
We can define two different equivalence
relations on $\sL$. With this aim in mind let us take two line operators $L_1$ and $L_2$ in $\sL$. The first relation, which we denote ``$\sim$'',
asserts that $L_1\sim L_2$ iff there exists a line changing
(0-dimensional) operator between them. This is the equivalence relation used when determining
which line operators survive screening, and
\be\label{LOL}
\widehat{\Gamma}^{(1)}\df\sL/{\sim} \,, \qquad   L_1 \sim L_2 \Leftrightarrow \ \exists \text{ local operator $\mathcal{O}$ at junction between  $L_1$ and $L_2$} \,,
\ee
is the group of lines charged under the 1-form symmetry. Its Pontryagin dual $\Gamma^{(1)}\df\Hom(\widehat{\Gamma}^{(1)}, U(1))$ is the group of 1-form
symmetries. We can also impose a finer equivalence relation, denoted $\sim'$,
which asserts that $L_1\sim' L_2$ if there exists a line changing
operator, transforming in a representation of $\cF^{(0)}$, between $L_1$ and
$L_2$. We denote the resulting group by 
\[
  \label{cEDef}
  \widehat{\cE}\df \sL/{\sim'} \,.
\]
There is a surjective map
$\alpha\colon \widehat{\cE}\to \widehat{\Gamma}^{(1)}$, since in our
definition of $\widehat{\Gamma}^{(1)}$ we did not impose that the line
changing operator is in a representation of $\cF^{(0)}$, it could be
in a representation of $F$ that does not descend to a representation
of $\cF^{(0)}$. The kernel of this map, $\ker \alpha$, is a subgroup
of $\widehat{C}$, where the hat indicates Pontryagin duality:
$\widehat{G} \df \Hom(G, U(1))$ for any abelian group $G$.\footnote{\,
  In our case both $\Gamma^{(1)}$ and $C$ will be finite abelian
  groups, so Pontryagin duality gives back the same group, but it is
  useful to keep the hats on to distinguish between the charged
  objects (wearing hats) and the associated symmetries (without
  hats).}  In all the examples in this paper
$\ker\alpha=\widehat{C}$. Physically, this group can be understood as
the group of line operators ending on point operators charged under
$C$.

Proceeding in this way, one obtains a short exact sequence of abelian
groups
\be
  \label{eq:dual-ses}
  0 \to \widehat{C} \longrightarrow \widehat{\cE} \stackrel{\alpha}{\longrightarrow} \widehat{\Gamma}^{(1)} \to 0
  \, ,
\ee
where we have set $\ker\alpha = \widehat{C}$. Our main goal in this
paper will be to reformulate this exact sequence in terms of the
geometry of the link.

It is convenient to dualise~\eqref{eq:dual-ses}, in which case we have
the short exact sequence 
\[
  \label{eq:cohom-extension}
  0 \to \Gamma^{(1)} \to \cE \to C \to 0\, .
\]
The non-trivial extensions are characterized by
$\Ext(C, \Gamma^{(1)})$.  In the following we will refer to the group
$\mathcal{E}$ as the {\it group extension of $C$ by $\Gamma^{(1)}$}. A
given element of $\Ext(C, \Gamma^{(1)})$ determines a Bockstein map
\[
\Bock\colon \quad H^n(\text{--};C)\to H^{n+1}(\text{--};\Gamma^{(1)})
\]
for the associated long exact sequence in cohomology. Note that
$\Bock$ is a cohomology operation \cite{Hatcher}, and is therefore an
element of $H^{n+1}(K(C,n); \Gamma^{(1)})$, where $K(C, n)$ is the $n$'th Eilenberg-MacLane space for the group $C$. We can show that this
group is indeed isomorphic to $\Ext(C, \Gamma^{(1)})$ as follows. The
case of interest to us is $n=2$, but we include a proof valid for
$n>1$. (The $n=1$ case is standard.) By definition $\pi_{i}(K(C,n))=C$
for $i=n$ and zero otherwise. By the Hurewicz theorem
$h\colon \pi_{n+1}(K(C,n))\to H_{n+1}(K(C,n))$ is
surjective for $n>1$. Since $\pi_{n+1}(K(C,n))=0$ we have
$H_{n+1}(K(C,n))=0$, and the existence of an isomorphism
$i\colon \Ext(C, \Gamma^{(1)})\to H^{n+1}(K(C, n), \Gamma^{{(1)}})$
then follows from the universal coefficient theorem.

We note that in the cases of interest to us in this paper we have
$C=\bZ_2$ and $\Gamma^{(1)}=\bZ_n$, and
$\Ext(C, \Gamma^{(1)})=\Ext(\bZ_2,\bZ_n)=\bZ_n/2\bZ_n=\bZ_{\gcd(2,n)}$
\cite{Hatcher}. So if $n$ is odd there is no non-trivial extension,
and therefore no non-trivial 2-group, while if $n=2k$ we have
$\Ext(\bZ_2,\bZ_{2k})=\bZ_2$. The non-trivial Bockstein operation in
this case is
$\Sq1\colon H^{n}(\text{--};\bZ_2)\to H^{n+1}(\text{--};\bZ_2)$
composed with the operation
$H^n(\text{--};\bZ_2)\to H^{n}(\text{--};\bZ_{2p})$ induced by the
non-trivial $\bZ_2\to \bZ_{2p}$ homomorphism (which is simply
multiplication by $p$).

\paragraph{2-Groups.}
The finite group $C$ also participates on a second short exact sequence
\be
  0 \to C \to F  \to  \mathcal{F}^{(0)} \to 0 \, .
\ee
This is a central extension of $\cF^{(0)}$ by $C$, with associated
characteristic class $w_2\in H^2(\cF^{(0)}; C)$. The non-triviality of the
2-group is then measured by the class
\[
\Bock(w_2)\in H^3(\cF^{(0)};\Gamma^{(1)}) \,.
\]
In the cases of interest to us, both $w_2$ and
$H^3(\cF^{(0)};\Gamma^{(1)})$ are non-trivial, and we can compute the
value of $\Bock(w_2)$ using the explicit characterisation in terms of
$\Sq1$ given above. As an example, consider the case $\cF^{(0)}=SO(3)$
and $C=\Gamma^{(1)}=\bZ_2$. If the short exact
sequence~\eqref{eq:cohom-extension} is non-trivial $\Bock$ is
non-trivial as a cohomology operation. So $\Bock=\Sq1$, since this is
the only non-trivial cohomology operation. On $BSO(3)$ we have, from
the Wu formula, $\Bock(w_2)=\Sq1(w_2)=w_3$, so we have a non-trivial
2-group.

In this paper we will first of all determine 
whether~\eqref{eq:cohom-extension} splits or not by computing $\widehat{\mathcal{E}}$. From this we can infer also the global form of
the flavor symmetry group by taking the quotient
\[
\mathcal{F}^{(0)} = {F\over C} \,.
\]
 2-groups of this type
were determined for 5d SCFTs with single gauge factors in
\cite{Apruzzi:2021vcu}, in 4d $\cN=1$ gauge theories in
\cite{Lee:2021crt} and in 6d (and more generally for gauge theories
with matter in any dimension $d=3,\cdots, 6$) in
\cite{Apruzzi:2021mlh}.

\subsection{2-Group Symmetries from Link Topology}\label{sec:core}

We now give a geometric realization of~\eqref{eq:dual-ses}. 
We will
focus on field theories arising from M-theory on singular cones
$\mathcal X^{d+1}$ with link $\textbf{L}^d$, geometrically engineering a
$(10-d)$-dimensional field theory $\cT_{\mathcal X}$.

We focus on the case in which the singularity of $\mathcal X^{d+1}$ is not
isolated and the non-compact loci supporting the corresponding non-isolated singularities are of dimension $d-3$. From the point of view of the field theory $\cT_{\mathcal X}$ the gauge
bosons living on the non-compact singular locus in this setting lead to a flavor
symmetry. In this paper we assume that the flavor symmetry of the theory $\cT_{\mathcal X}$ is faithfully reproduced by the geometry of these loci. 
The non-compact locus of the singularity will intersect
$\mathbf{L}^d$ along a subvariety $\mathcal S_0$. We denote by $\mathcal S$ a small tubular
neighbourhood of $\mathcal S_0$ inside $\mathbf{L}^d$.

We want to understand the short exact sequence~\eqref{eq:dual-ses}
from the geometric viewpoint. The geometric interpretation of the
group $\widehat{\Gamma}^{(1)}$ is by now standard, and has been studied e.g. in \cite{Morrison:2020ool,Albertini:2020mdx}. The lines in $\widehat{\Gamma}^{(1)}$
charged non-trivially under 1-form symmetries arise from M2 branes
wrapping non-compact 2-cycles which intersect $\mathbf{L}^d$ along
representative cycles of non-trivial elements of $H_1(\mathbf{L}^d)$. This
group is purely torsional in the cases of interest to us.

\begin{figure}
  \centering
  \includegraphics[height=4cm]{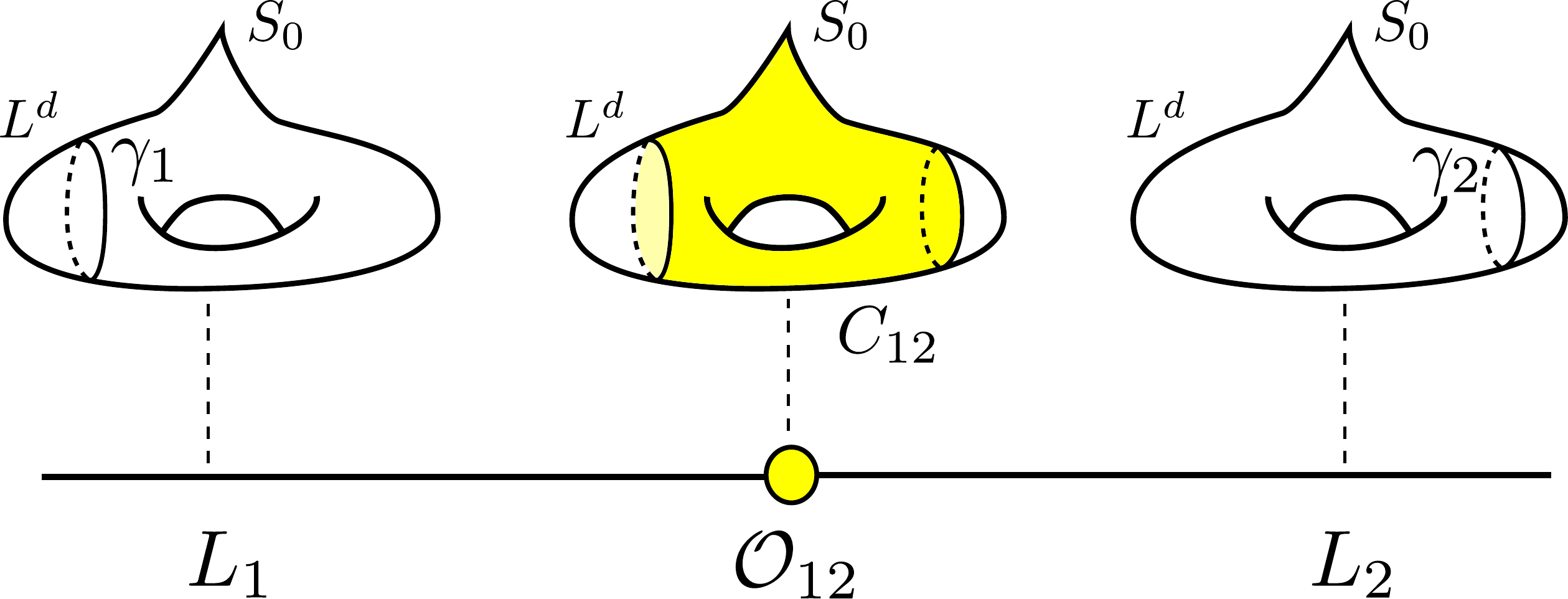}
  \caption{The geometric description of a line changing operator $\cO_{12}$ connecting line operators $L_1$ and $L_2$ as a chain  (shown in yellow) connecting homologous cycles on the boundary. }
  \label{fig:tilde-geometry}
\end{figure}

It will be illuminating to give an interpretation of this familiar
result in terms of line-changing operators, and the equivalence
relation $\sim$ defined above in (\ref{LOL}). Consider two M2 branes wrapping
non-compact cycles $\Sigma_1$ and $\Sigma_2$, giving rise to line
operators $L_1$ and $L_2$. These cycles will intersect $\mathbf{L}^d$ on curves
$\gamma_1$ and $\gamma_2$. Assume that these two curves are in the
same homology class, so there is a chain $C_{12}$ such that
$\partial C_{12}=\gamma_1 - \gamma_2$. In this case we can construct a
line-changing operator $\cO_{12}$ between $L_1$ and $L_2$ by wrapping
an M2 brane on the chain $C_{12}$ (times the radial direction). This
is shown in figure~\ref{fig:tilde-geometry}.

This interpretation of the relations in $\sim$ immediately leads to an
interpretation of $\widehat{\cE}$, and thereby the two-group symmetry. Recall that in this case we want to
quotient the space of lines by $\sim'$ in (\ref{cEDef}), which does not include
line-changing operators charged under $C$. We can accomplish this from
the point of view of the geometry by excising $\mathcal S$ from $\mathbf{L}^d$, we
denote the resulting space $\mathbf{L}^d-\mathcal S$. The chains that pass through the
singularity get an extra boundary after excising $\mathcal S$, \textit{and therefore no
longer lead to relations $\gamma_1=\gamma_2$ in homology}. We will
argue below that this extra boundary encodes the charge under $C$. The
surviving homological relations therefore come from chains in
$\mathbf{L}^d - \mathcal S$, and are precisely those uncharged under the centre of the
flavor group. So we identify 
\be
\widehat{\cE}=H_1(\mathbf{L}^d- \mathcal S) \,.
\ee

\begin{figure}
  \centering
  \includegraphics[height=4cm]{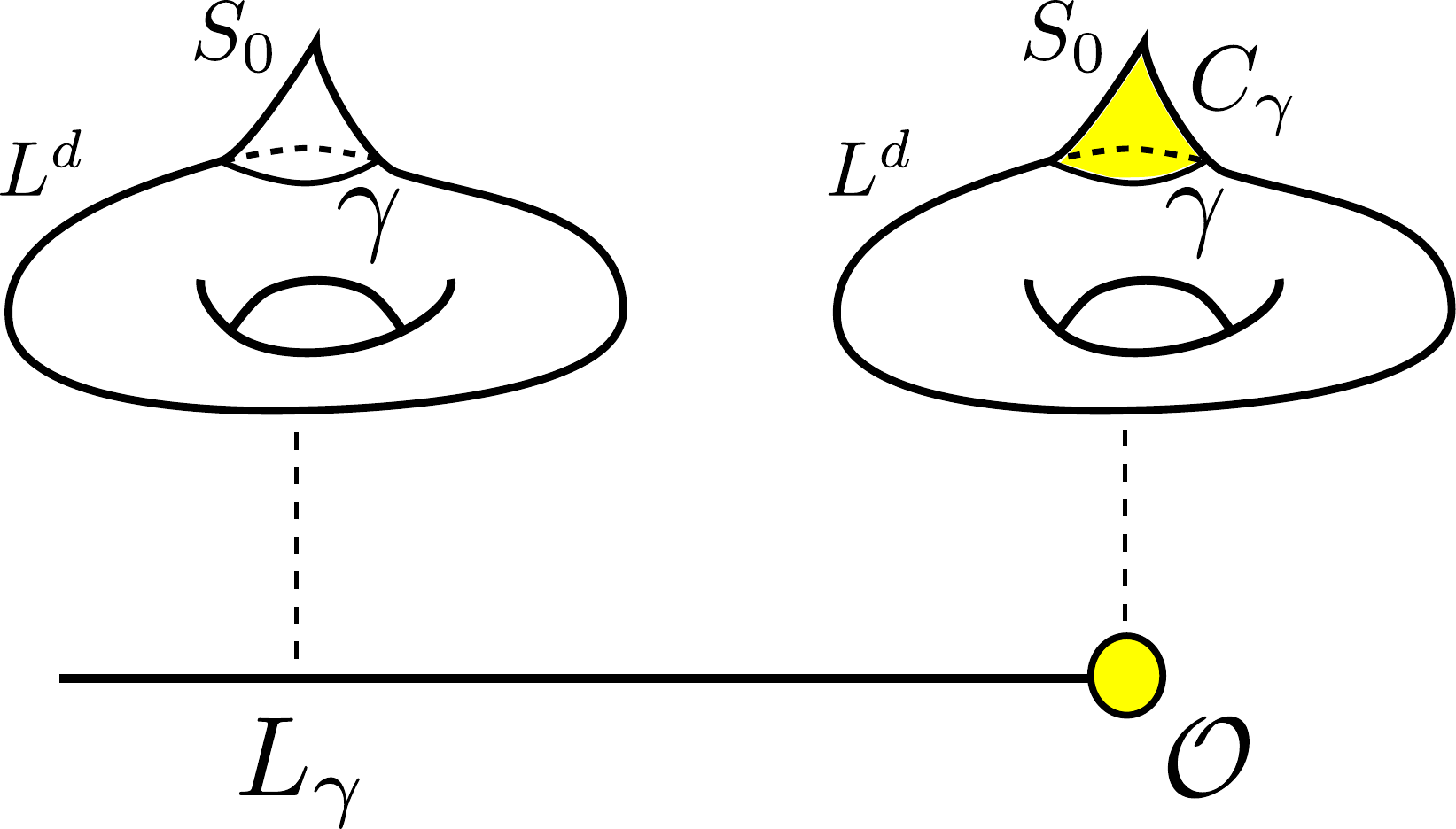}
  \caption{A line operator associated to a homologically trivial curve
    $\gamma$ on $\mathbf{L}^d$ ending on a point operator. From the point of
    view of the internal geometry $\mathbf{L}^d$ the point operator at the end
    corresponds to the M2-brane worldvolume wrapping a chain $C_\gamma$ with
    boundary $\gamma$.}
  \label{fig:endable-line}
\end{figure}

Finally, we can give an interpretation of $\widehat{C}$
in~\eqref{eq:dual-ses} along similar lines. Consider a non-compact M2
brane intersecting $\mathbf{L}^d$ along a closed curve $\gamma$ belonging to a
non-trivial homology class on $\mathbf{L}^d- \mathcal S$ that becomes trivial when
embedded in $\mathbf{L}^d$. The associated line operator $L_\gamma$ is
therefore trivial as an element of $\widehat{\Gamma}^{(1)}$, because the
line can end on a point operator $\cO$, given by an M2-brane wrapping a
chain $C_{\gamma}$ with boundary $\gamma$ --- see
figure~\ref{fig:endable-line}. These chains will be non-trivial under
$\sim'$ if $C_\gamma$ is charged under $C$, which we can detect by
computing the class of $\gamma$ in $\Tor H_1(\partial \mathcal S)$. This
mirrors the standard computation of the charge of a line operator
under the 1-form symmetries of the field theory summarised above, with
the difference that we are viewing the M2 brane wrapped on $C_\gamma$
as a Wilson line for the seven dimensional gauge theory living on the non-compact singular locus (which from the point of view of the 5d SCFT is a flavor sector).

Collecting the results of our discussion so far, we have
translated~\eqref{eq:dual-ses} into the geometric statement that the
following sequence is exact:
\[
  \label{eq:geometric-ses}
  0 \to \Tor H_1(\partial \mathcal S) \to H_1(\mathbf{L}^d - \mathcal S) \to H_1(\mathbf{L}^d) \to 0\, .
\]
Although we have derived this exact sequence by reinterpreting the field theory discussion geometrically,
it is also possible to derive it via purely geometric arguments as
follows. Assume that we have a space $X$, and two subspaces
$A,B\subset X$ such that the union of their interiors covers $X$. Then
there is a long exact sequence known as the Mayer-Vietoris spectral
sequence \cite{Hatcher} that reads
\[
  \ldots \to \tilde H_2(X) \to \tilde H_1(A\cap B) \to \tilde
  H_1(A)\oplus \tilde H_1(B) \to \tilde H_1(X) \to \tilde H_0(A\cap B) \to
  \ldots
\]
where the tildes denote reduced homology groups. For simplicity, in
our analysis we will assume that $A$, $B$, $X$ and $A\cap B$ are all
connected. (The generalisation is straightforward, but a little
cumbersome.) Additionally, we will assume that the boundary map
$\tilde H_2(X) \to \tilde H_1(A\cap B)$ vanishes. We do not have a
general argument for this, but it is possible to verify in our
explicit examples below that it is the case. With these assumptions
in place, the Mayer-Vietoris long exact sequence implies the short
exact sequence
\[
  0 \to H_1(A\cap B) \to H_1(A)\oplus H_1(B) \to H_1(X) \to 0\, .
\]
We can obtain~\eqref{eq:geometric-ses} from here by taking
$X=\mathbf{L}^d$, $A=\mathbf{L}^d-\mathcal S$ and $B= \mathcal S'$,
where $\mathcal S'$ is a slight thickening of the tubular
neighbourhood $\mathcal S$, so that the interiors of
$\mathbf{L}^d-\mathcal S$ and $\mathcal S'$ indeed cover
$\mathbf{L}^d$. The last term clearly is as
in~\eqref{eq:geometric-ses} under this substitution. To recover the
other two terms we start by noting that in the geometries analysed
below $\mathcal S=T\times S^1$, so
$\partial \mathcal S = \partial T \times S^1$, for some singular toric
cone $T$ with boundary $\partial T$ (For instance, in some of the
examples below we will have $T$ a neighbourhood of the singular point
in $\bC^2/\bZ_n$, and therefore $\partial T=S^3/\bZ_n$, although we
emphasise that our analysis is more general). The space
$(\mathbf{L}^d-\mathcal S)\cap \mathcal S'$ deformation retracts to
$\partial S=\partial T\times S^1$, so the first term becomes
$H_1(\partial T)\oplus \bZ$. Singular toric cones of complex dimension
1 and 2 have $H_1(\partial T)$ purely torsional, so we can
equivalently write $H_1(\partial S)=\bZ\oplus \Tor H_1(\partial
T)$. For the middle term, we have
$H_1(B)=H_1(\mathcal S')=H_1(\mathcal S)=H_1(T)\oplus H_1(S^1)=\bZ$,
using that toric varieties have no non-trivial 1-cycles. An explicit
analysis of the inclusion map $H_1(A\cap B) \to H_1(A)\oplus H_1(B)$
then shows that it restricts to an isomorphism on the $\bZ$ factors
comings from the $S^1$ factor in $\cS$, so we
recover~\eqref{eq:geometric-ses} also as a mathematical consequence of
Mayer-Vietoris.

One technical point to highlight
at this stage is that in the derivation above we have used rather
special properties of the geometries studied in this paper to conclude
that the Mayer-Vietoris long exact sequence splits, so that we end up
with a short exact sequence for the groups of interest. In particular,
our examples below are such that $\partial T$ is a lens space
$S^3/\bZ_n$, associated with a flavor algebra $\fsu(n)$, and our
analysis gives $C=\bZ_n$, which agrees with what one finds from field
theory considerations in these cases. In more general situations we do
not expect the Mayer-Vietoris sequence to split, but we can still
write a tautological short exact sequence
\[
  0 \to \ker(a) \to H_1(\bfL^d-\cS)\oplus H_1(\cS') \xrightarrow{a}
  H_1(\bfL^d) \to 0\, ,
\]
still under the assumption that the $\partial \cS$ is connected. In
this case we would identify $C=\Tor\ker(a)$.

\medskip

Our main task therefore becomes to compute 
\be
\widehat{\cE}=H_1(\mathbf{L}^d -\mathcal S).
\ee
Below we will introduce methods that allow us to compute this group systematically in ordinary and generalised toric varieties, but before
doing so let us comment briefly on how this discussion connects to
previous work \cite{Apruzzi:2021vcu,Apruzzi:2021mlh}. Consider the long exact sequence for the pair
$(\mathcal X_\epsilon^{d+1}, \mathbf{L}^d - \mathcal S)$, where $\mathcal X_\epsilon^{d+1}$ is a neighborhood of
the origin of the conical singularity $\mathcal X^{d+1}$ (so that $\partial \mathcal X_\epsilon^{d+1}=\mathbf{L}^d$):
\[
  \ldots \to H_2(\mathcal X_\epsilon^{d+1}) \to H_2(\mathcal X_\epsilon^{d+1}, \mathbf{L}^d - \mathcal S) \to H_1(\mathbf{L}^d -
  \mathcal S) \to H_1(\mathcal X_\epsilon^{d+1})\to \ldots
\]
Since $\mathcal X_\epsilon^{d+1}$ is a special holonomy variety we
expect $H_1(\mathcal X_\epsilon^{d+1})=0$ (this holds for toric
varieties and singular hypersurfaces, for instance; we assume that this
is the case for the validity of this analysis), and the sequence
terminates:
\[
  \label{eq:boundary-bulk}
  \ldots \to H_2(\mathcal X_\epsilon^{d+1}) \xrightarrow{i} H_2(\mathcal X_\epsilon^{d+1}, \mathbf{L}^d -
  \mathcal S) \to H_1(\mathbf{L}^d - \mathcal S) \to 0
\]
which implies
\[
  \widehat{\cE} = H_1(\mathbf{L}^d - \mathcal S) = \coker(i) \, .
\]
This statement can be interpreted in terms of screening, generalising
the discussion in \cite{DelZotto:2015isa,GarciaEtxebarria:2019caf} to
\mbox{2-groups}: the lines in
$H_2(\mathcal X_\epsilon^{d+1}, \mathbf{L}^d - \mathcal S)$ are lines
where we keep track of the flavor charge --- the fact that we are
excising $\mathcal S$ from $\mathbf{L}^d$ in the pair means that
relative cycles with different flavor charge, that would be
equivalent in $H_2(\mathcal X_\epsilon^{d+1}, \mathbf{L}^d)$, are no
longer equivalent in
$H_2(\mathcal X_\epsilon^{d+1}, \mathbf{L}^d - \mathcal S)$, since the
chain connecting them does necessarily pass through $\mathcal S$ (as
in our boundary analysis above).

%%%%%%%%%%%%%%%%%%%%%%%%%%%%
\section{2-Group Symmetries in 5d from M-theory}\label{sec:5dcase}

In this section we explain how to exploit the general method discussed in the previous section to recover the 
results on 2-groups from the geometry of the boundary in the context of 5d SCFTs arising 
from compactification of M-theory on local CY singularities $\mathcal X$. For simplicity, we will mostly focus on cases where
$\mathcal X$ is a toric CY cone, since in these cases we can apply results from  \cite{Garcia-Etxebarria:2016bpb} to capture the 
geometry of the boundary. For toric CY singularities the corresponding 5d SCFTs also have a dual geometric engineering 
in terms of webs of $(p,q)$ five-branes --- this gives rise to a dual IIB version of the excision method which is extremely powerful in practice (see section \ref{sec:Webs}).

In particular, in section  \ref{sec:yay!} we show the equivalence of the methods presented in this paper with the prescription for determining the 2-group structure uncovered in \cite{Apruzzi:2021vcu} for the case of CY threefolds $\mathcal X$ whose non-compact singularities faithfully reproduce the flavor symmetries of the corresponding 5d SCFTs.

Our analysis in this section proves and extends a proposal in
\cite{DelZotto:2022fnw}. The authors of that paper consider orbifolds
of the form $S^5/\Gamma$ with a normal subgroup $H\triangleleft\Gamma$
acting with fixed points on $S^5$, and proposed that in these cases
the 2-group structure is associated with the short exact sequence
\[
  0 \to H_1(S^5/\Gamma) \to \Gamma^{\text{ab}} \to \Gamma^{\text{ab}}/H_1(S^5/\Gamma) \to 0
\]
where $\Gamma^{\text{ab}}\df \Gamma/[\Gamma,\Gamma]$ denotes the
abelianisation of $\Gamma$. We can reinterpret this in terms of our
discussion above: this sequence is the Pontryagin dual
of~\eqref{eq:geometric-ses}, noticing that
$\pi_1(S^5/\Gamma - S)=\Gamma$ (since $\Gamma$ acts freely on
$S^5/\Gamma - S$), and by the Hurewicz isomorphism
$H_1(S^5/\Gamma - S)=\pi_1(S^5/\Gamma-S)^{\text{ab}}$.

\medskip

In the section \ref{sec:beyondtoric} below we discuss a possible generalization of our arguments for SCFTs that arise outside of the toric realm.

\subsection{2-groups from the Boundary: the case of Toric 5d SCFT}

\begin{figure}

\begin{center}
  \begin{minipage}{2cm}
    $\begin{gathered}\xymatrix@=0.5em{&\bullet_{\nu-1}\ar@{-}[r]\ar@{-}[dl]&\cdots\\ \bullet_\nu\ar@{-}[d]&&&&\vdots\\ \bullet_1\ar@{-}[r]&\bullet_2\ar@{-}[r]&\bullet_3\ar@{-}[r]&\bullet_4\ar@{-}[ur]}\end{gathered}$
    \vspace{3cm}
  \end{minipage}
  \hspace{3cm} \includegraphics[scale=0.6]{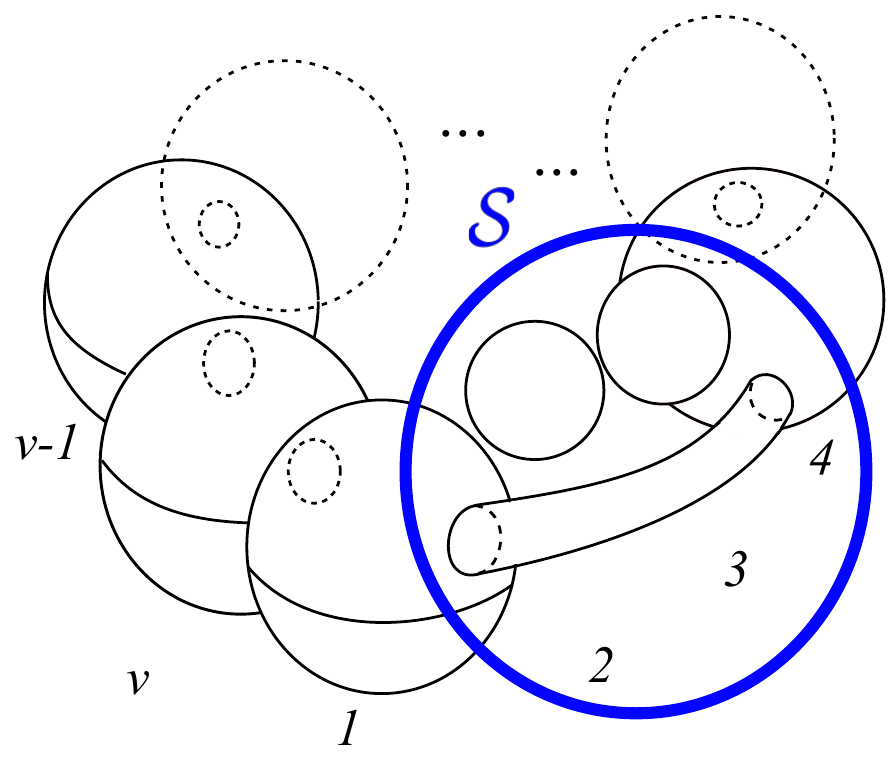}
  \vspace{-3cm}
\end{center}

\caption{Schematic topology of $\mathcal C_{\mathbf{L}_{\mathcal X}}^3$ \cite{Garcia-Etxebarria:2016bpb} -- see also \cite{Albertini:2020mdx} for the case $\mathcal X$ non-isolated. In this figure we have 4 vertices along an external edge corresponding to a $\mathbb C^2/\mathbb Z_3$ singularity. We denote as $\mathcal S$ the neighborhood of the singular locus we need excise when considering the 2-group structure -- clearly after the excision the Lens spaces $\mathbf{L}_1$ and $\mathbf{L}_4$ in the above figure become contratible and do not contibute to $H_1(\mathbf{L}^5_{\mathcal X} - \mathcal S)$.}\label{fig:superballs}
\end{figure}

Consider the case $\mathcal X$ is a toric CY singularity. 
We will first apply our general singularity excision approach to two-groups in this context and show subsequently the equivalence with the direct intersection computation. 

To each such singularity corresponds a toric diagram, a convex polytope embedded in $\mathbb Z^2$ with $\nu$ external vertices $\mathbf{v}_i \in \mathbb \mathbb \mathbb Z^2$, $i=1,...,\nu$. In this paper we are interested in the geometry of the link of $\mathcal X$, which we denote $\mathbf{L}_{\mathcal X}^5$. As first argued (to the best of our knowledge) in \cite{Garcia-Etxebarria:2016bpb}, one has that
\begin{equation}
  H_{n}(\mathbf{L}_{\mathcal X}^5) = H_n(\mathcal B_{\mathbf{L}_{\mathcal X}}^3) \qquad \text{for } n \leq 2\, ,
\end{equation}
where $\mathcal B_{\mathbf{L}_{\mathcal X}}^3$ is a 3-chain of lens spaces
\be
\mathcal B_{\mathbf{L}_{\mathcal X}}^3 \simeq \mathbf{L}_{n_1} \veebar \mathbf{L}_{n_2}  \veebar  \ldots \veebar \mathbf{L}_{n_\nu}
\ee
where $\veebar$ denotes that the lens spaces are joined along their torsion cycle. The $n_i$ are determined as follows. For each
external vertex $\mathbf{v}_i$, $i\in\{1,\ldots,\nu\}$, construct the triangle
$T_i$ defined by the vertex and the two vertices adjacent to it, that
is, the convex hull of $\{\mathbf{v}_{i-1},\mathbf{v}_i,\mathbf{v}_{i+1}\}$ (with $\mathbf{v}_0\df \mathbf{v}_\nu$
and $\mathbf{v}_{\nu+1}\df \mathbf{v}_1$). Then 
\be
n_i=2\mathrm{Area}(T_i)\,.
\ee
Whenever we have a collection of $m+1$ external lattice points $\mathbf{v}_i$ along an edge, this corresponds to the presence of a non-compact curve of singularities $\mathbb C^2/\mathbb Z_m$, 
giving a factor $\mathfrak{su}(m)$ of the global symmetry of the 5d SCFT $\mathcal T_{\mathcal X}$. If this is the case, the corresponding triangles $T_i$ will
have zero area. Since the lens space $\mathbf{L}_n$ is a circle fibration over $\mathbf{S}^2$ of degree $n$ ($\mathbf{L}_n \simeq S^3/\mathbb Z_n$ for $n\geq 1$) we can include the case $n=0$ as $\mathbf{L}_0\cong \mathbf{S}^2\times \mathbf{S}^1$ --- whenever we have points along an edge, upon crepant resolution the local geometry of the $\mathbf{T}^2$ fiber considered in
\cite{Garcia-Etxebarria:2016bpb} around the point is that of
$\mathbf{S}^2\times \mathbf{S}^1$ (see figure \ref{fig:superballs}). Additionally,
one can show that \cite{Garcia-Etxebarria:2016bpb}
\begin{equation}
  H_1(\mathcal B_{\mathbf{L}_{\mathcal X}}^3) = \mathbb Z_{\gcd(n_1,\ldots,n_\nu)}\, ,
\end{equation}
and in the toric case
\begin{equation}
  \label{eq:toric-torsion}
  \Gamma^{(1)} \simeq H_1(\mathbf{L}_{\mathcal X}^5) = \mathbb Z_{\gcd(n_1,\ldots,n_\nu)}
\end{equation}
taking into account that $\gcd(0,\ldots)=\gcd(\ldots)$.

\medskip

Now consider the case we have a single non-compact curve of singularities, corresponding to the fact that the flavor symmetry of the 5d SCFT has Lie algebra $\mathfrak{su}(m)$. As discussed above this corresponds to a sequence of outer vertices $\mathbf{v}_{i_1}$, $\mathbf{v}_{i_2}$, $\ldots \mathbf{v}_{i_{m+1}}$ which are all aligned along an external edge. Then we can explicitly apply the method outlined in section \ref{sec:generalcase} to recover the 2-group structure for the 5d SCFT at hand. In presence of such a singularity, the link itself will have a singular locus $\mathcal S$ and we are interested in computing
\be
\hat{\cE} = H_1(\mathbf{L}_{\mathcal X}^5 - \mathcal S) \,.
\ee
The latter is easily obtained from the discussion in \cite{Garcia-Etxebarria:2016bpb}. Removing the neighborhood of the singular locus $\mathcal S$ alters the topology of the 3-chain $\mathcal B_{\mathbf{L}_{\mathcal X}}^3$, rendering contractible the Lens spaces $\mathbf{L}_{v_{i_1}}$ and $\mathbf{L}_{v_{i_{m+1}}}$. We denote the resulting 3-chain $\widehat{\mathcal B}_{\mathbf{L}_{\mathcal X}}^3$. We can always relabel the outer vertices so that the first $m+1$ corresponds to the $m+1$ aligned ones. Proceeding in this way, we obtain that
\be
\widehat{\mathcal B}_{\mathbf{L}_{\mathcal X}}^3 \sim \mathbf{L}_{m+2} \veebar \mathbf{L}_{m+3} \veebar \cdots \veebar \mathbf{L}_{\nu}
\ee 
where $\sim$ is homotopy equivalence. Then by the same argument that lead to the conclusion in \cite{Garcia-Etxebarria:2016bpb}, we obtain that
\be
\hat{\cE} = H_1(\mathbf{L}_{\mathcal X}^5 - \mathcal S) = H_1(\widehat{\mathcal B}_{\mathbf{L}_{\mathcal X}}^3) = \mathbb Z_{\gcd(n_{m+2},\ldots,n_\nu)}\,.
\ee

\paragraph{Example: $SU(2)_0$.}
Let us consider the simplest example: the 2-group of the 5d SCFT, which has a Coulomb branch description as $SU(2)_0$ \cite{Apruzzi:2021vcu}. 
The toric diagram is 
\be
\begin{tikzpicture}[x=1cm,y=1cm]
\draw[step=1cm,gray,very thin] (-1,0) grid (1,2);
\draw[ligne] (-1,0)--(0,0)--(1,0)--(0,2)--(-1,0); 
%\draw[lignet] (-1,0) -- (0,1) -- (0,0) ;
%\draw[lignet] (1,0) -- (0,1) -- (0,2) ;
\node[bd] at (-1,0)  [label = below: \large{$1$}] {};
\node[bd] at (0,0) [label = below: \large{$2$}] {};
\node[bd] at (1,0)  [label = below: \large{$3$}] {};
\node[bd] at (0,2) [label = above: \large{$4$}] {};
\node[bd] at (0,1) {};
\end{tikzpicture} 
\ee
where we labeled the external vertices. 
Now we apply the above algorithm to compute the 1-form symmetry by considering the triangles $T_i \df \Delta(i-1,i,i+1)$: 
\be
\begin{tikzpicture}[x=1cm,y=1cm]
\draw[step=1cm,gray,very thin] (-1,0) grid (1,2);
\draw[ligne] (-1,0)--(0,0)--(1,0)--(0,2)--(-1,0); 
\draw[lignet] (-1,0) -- (0,1) -- (0,0) ;
\draw[lignet] (1,0) -- (0,1) -- (0,2) ;
\node[bd] at (-1,0) [label=below: \large{$\mathbf{S}^3/\mathbb{Z}_{2}$}] {};
\node[bd] at (0,0) {}; 
\node[bd] at (1,0)  [label=below: \large{$\mathbf{S}^3/\mathbb{Z}_{2}$}] {};
\node[bd] at (0,2)  [label=above: \large{$\mathbf{S}^3/\mathbb{Z}_{4}$}] {};
\node[bd] at (0,1) {};
\end{tikzpicture} 
\ee
In orange we show the triangulations. E.g. for the vertex 3 we find $|\Delta (2,3,4)| = n_3= 2$ etc. 
Thus the 1-form symmetry is $\Gamma^{(1)}= \mathbb{Z}_{\gcd (2,2,4)} = \mathbb{Z}_2$, as expected for this theory. 

To determine $\widehat{\mathcal{E}}$ requires the excision of the flavor nodes. In this example the only edge is the bottom edge, which has the vertex $2$ on it. Thus we find 
\be
\begin{tikzpicture}[x=1cm,y=1cm]
\draw[step=1cm,gray,very thin] (-1,0) grid (1,2);
\draw[ligne] (-1,0)--(0,0)--(1,0)--(0,2)--(-1,0); 
\draw[lignet] (-1,0) -- (0,1) -- (0,0) ;
\draw[lignet] (1,0) -- (0,1) -- (0,2) ;
\node[bd] at (-1,0) {};
\node[fd] at (0,0) {}; 
\node[bd] at (1,0)  {};
\node[bd] at (0,2)  [label=above: \large{$\mathbf{S}^3/\mathbb{Z}_{4}$}] {};
\node[bd] at (0,1) {};
\end{tikzpicture} 
\ee
where the green vertex is the one we excised. The remaining asymptotic topology is $\mathbf{S}^3/\mathbb{Z}_4$ and thus 
\be
\widehat{\mathcal{E}} = \mathbb{Z}_4
\ee
consistent with the existence of the 2-group \cite{Apruzzi:2021vcu}. Out of this analysis we also infer the global form of the flavor symmetry for this theory is $SU(2)/\mathbb{Z}_2$, consistent with \cite{Kim:2012gu, BenettiGenolini:2020doj}. 

\medskip

The above discussion can be generalized to the cases where the singularity $\mathcal X$ has several non-compact singularities, each corresponding to a factor $\mathfrak{su}(m_k)$ of the flavor symmetry. It is natural to consider each of these factors separately, excising only the neighborhood $\mathcal S_k$ of the corresponding singular locus on $\mathbf{L}^5_{\mathcal X}$. The analysis is the same we discussed above, and each of these factors would end up forming a different extension with the one form symmetry. For an explicit example of this see section \ref{sec:quivers} below.

\paragraph{Summary of the Computational Approach.} To summarize we find the following computational description for the 1-form symmetry and 2-group for toric geometries. 
\begin{enumerate}
\item Compute the asymptotic lens space topology for each external vertex of the toric diagram, i.e. for each triple of vertices $v_{i-1}, v_i, v_{i+1}$ compute the volume of the triangle $\Delta (v_{i-1}, v_i, v_{i+1}) = n_i$, then the boundary topology is $S^3/\mathbb{Z}_{n_i}$.
The 1-form symmetry is then 
\be
\Gamma^{(1)} = \mathbb{Z}_{\gcd (n_1, \cdots, n_N)}. 
\ee 
where $N$ is the total number of external vertices. Note that vertices along edges get assigned $S^3/\mathbb{Z}_0 \sim S^2 \times S^1$. 
\item To determine $\mathcal{E}$ consider the vertices along edges. For each edge $e_\ell$, let 
\be
V_\ell  = \{v_i:  \ v_i \in e_\ell \ \text{and }\ v_i \cap (\partial e_\ell) = \emptyset \}
\ee 
be the vertices along the edge, however not including the corners. Note that these correspond to non-compact divisors, which generate a flavor symmetry algebra $\mathfrak{f} = \mathfrak{su} (|V_{\ell}| +1)$. 
For each edge we excise $V_\ell$ and $\partial e_\ell$ and compute 
\be
\mathcal{E}_{\ell} = \mathbb{Z}_{\gcd (\{n_i:\  v_i \not\in  e_\ell \}}\,,
\ee
i.e. we excise the lens spaces along the edge $e_\ell$ including the corners.
Then the associated flavor symmetry group is 
\be
\mathcal{F}_{\ell} = {SU (|V_{\ell}| +1) \over C_\ell} \,,\qquad C_{\ell}= {\mathcal{E}_{\ell} \over \Gamma^{(1)}} \,.
\ee
If this group  is a a non-trivial extension of  $\Gamma^{(1)}$
\be
1 \rightarrow \Gamma^{(1)} \rightarrow \mathcal{E}_\ell \rightarrow C_{\ell} \rightarrow 1 \,,
\ee
then there is a non-trivial 2-group if in addition there is  non-zero Postnikov class in 
\be
\Bock (w_2)= w_3 \in H^3 (B\mathcal{F}, \Gamma^{(1)}) \,.
\ee
Repeating this analysis along all edges $e_\ell$ results in the full symmetry structure of the theory, identifying which 0-form symmetry factors participate in the 2-group structure.  
\end{enumerate}

\subsection{Excision and $(p,q)$-Fivebrane Webs}\label{sec:Webs}

\begin{figure}
\begin{center}
\includegraphics[width= 0.65 \textwidth]{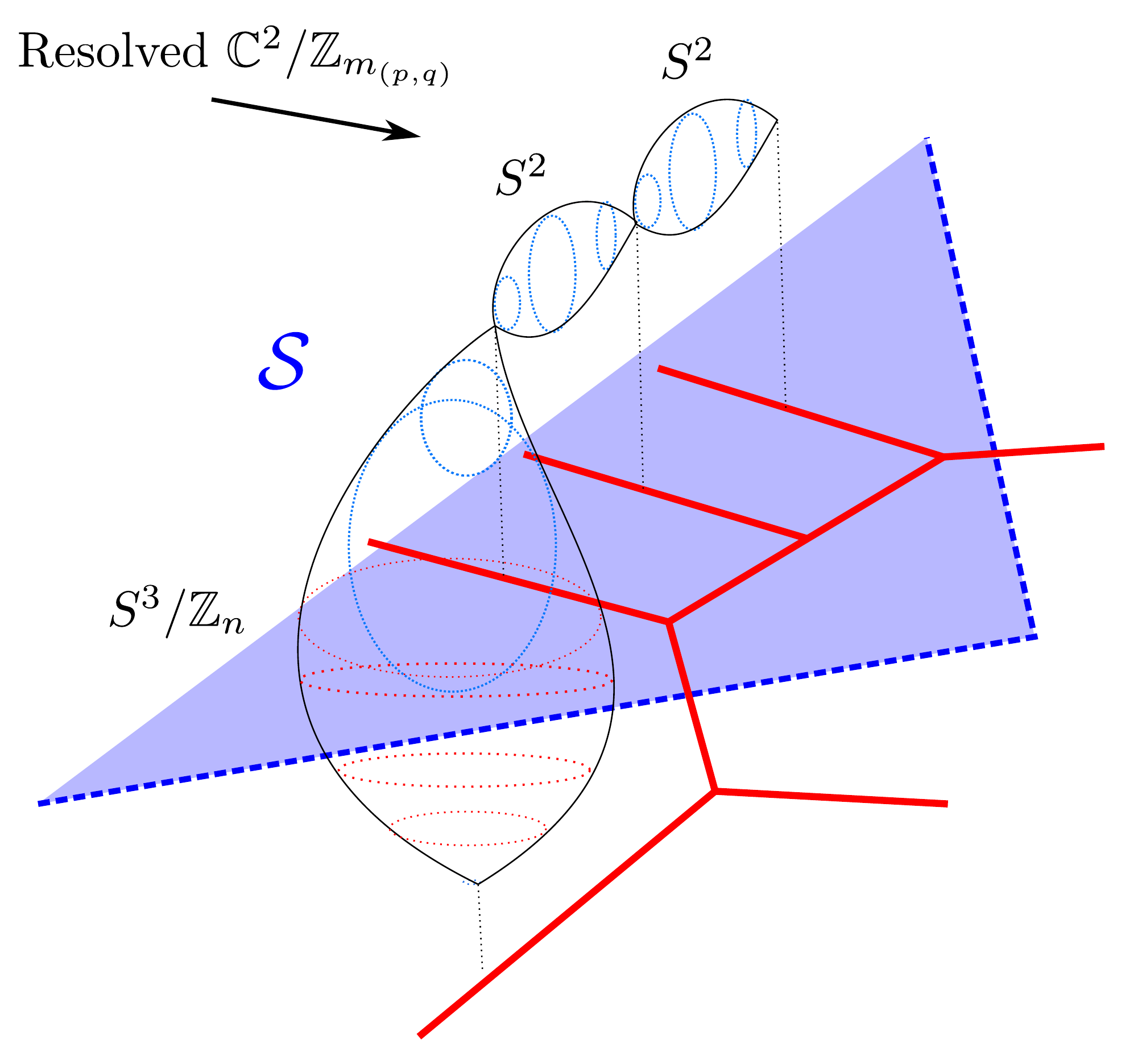}
\end{center}
\caption{Duality between geometry and $(p,q)$ web. The figure shows the combination of both brane-web and geometry. 
The location of the $(p,q)$ branes (in red) is dual to loci where $T^2$ cycles degenerate. This gives rise to the topologies we draw above. The excision locus, corresponding to the shaded region inside the wedge, is denoted by $\mathcal{S}$. Removing $\mathcal S$ is dual to deleting the corresponding parallel $(p,q)$ branes.}\label{fig:Alan}
\end{figure}

In the dual 5-brane webs the description becomes even simpler -- and combinatorially easy to implement. 
Consider a toric polygon for a Calabi-Yau three-fold, and let $W=\{(p_i, q_i)\}$ be the dual labels for the 5-brane web, i.e. the differences of consecutive external vertices in the polygon. The precise relation is that for a  $\mathbf{v}_i$ and $\mathbf{v}_{i+1}$ consecutive external vertices 
\be
\mathbf{v}_i- \mathbf{v}_{i+1}  = (a, b, 0)   \qquad \Leftrightarrow \qquad (p,q) = (b, -a) \,.
\ee
In this convention, the D5-branes are horizontal (i.e. charge $(1,0)$) and NS5 vertical (i.e. $(0,1)$). Notice that here it is key to include also vertices along a single edge (i.e. $(p,q)$ charges will have multiplicities). This duality between toric diagrams and $(p,q)$ webs has a beautiful gemetrical interpretation \cite{Aharony:1997bh,Leung:1997tw} -- see figure \ref{fig:Alan}. 
The diagram describing the positions of the $(p,q)$ fivebranes is dual in geometry to a fibration of a $T^2$ over a plane. The $(p,q)$ segments can be understood as characterising loci where a 1-cycle $pA + q B$ of the $T^2$ is shrinking to zero size. In presence of $m$ consecutive parallel (p,q) fivebranes extending at infinity we see that in the dual description we find a resolved ALE singularity of type $\mathbb C^2/\mathbb Z_m$ whose locus is a non-compact curve. On the contrary, the components of the boundary that are dual to regions between $(p,q)$ fivebranes that are non-parallel correspond to the lens-spaces, as the corresponding shrinking one-cycles change, giving rise to a Hopf-like fibration description of the space $S^3/\mathbb Z_n$. This explains why the topology of the boundary is encoded in the structure of the asymptotic (p,q) fivebranes that extend to infinity. It was indeed shown in \cite{Bhardwaj:2020phs}, that the 1-form symmetry can be computed from $W$ simply by taking the Smith normal form (SNF)
\be
\text{diag} (n_1, n_2) = \text{SNF} (W)\,.
\ee
To determine whether there is a non-trivial extension, we need to account for the flavor symmetry groups. As explained above, these are captured by the multiplicities in the $(p,q)$-charges of consecutive parallel $(p,q)$ fivebranes. Denote by 
\be
m_{(p,q)}= \text{multiplicity of }(p,q) \text{ in } W\,
\ee
one such multiplicity. In particular this means that the flavor symmetry algebra $\mathfrak{f}$ of the 5d SCFT has a subalgebra $\mathfrak{su}(m_{(p,q)})$.  Clearly now we see what is the dual procedure of the excision of the locus $\mathcal S$ corresponding to this singularity of $\mathbf{L}_{\mathcal X}^5$: this is dual to removing the corresponding collection of $m_{(p,q)}$ parallel $(p,q)$ webs corresponding to the $\mathbb C^2/\mathbb Z_{m_{(p,q)}}$ singularity --- see Figure \ref{fig:Alan}. Therefore we introduce 
\be
W^{\text{red}}_{(p,q)} = \text{Matrix obtained by removing }(p,q)^{m_{(p,q)}}  \text{ from } W \,.
\ee
The flavor symmetry subalgebra $\mathfrak{su}(m_{(p,q)})$ contributes a non-trivial extension to $\widehat{\mathcal{E}}$ if 
\be
\text{SNF} (W^{\text{red}}_{(p,q)} ) = \text{diag} (N_1, N_2) \,,
\ee 
where $N_i/n_i >1$ and gcd$(N_i/n_i, n_i) \not= 1$. 

\paragraph{Example: $SU(2)_0$.}
Let us apply this again to  the $SU(2)_0$ SCFT in 5d, which has the dual brane-web (shown in red): 
\be
\begin{tikzpicture}[x=1cm,y=1cm]
\draw[step=1cm,gray,very thin] (-1,0) grid (1,2);
\draw[ligne] (-1,0)--(0,0)--(1,0)--(0,2)--(-1,0); 
\draw[lignew] (-0.5,-1.5)-- (-0.5, 0.5) -- (0.5, 0.5) -- (0.5, -1.5);
%\draw[lignew] (-1.5, 1.5)-- (1.5, 1.5);
\draw[lignew]  (-0.5, 0.5) -- (-1.5, 1.5)-- (1.5, 1.5) -- (0.5, 0.5);
\draw[lignew]  (-1.5, 1.5) -- (-2.5, 1.75);
\draw[lignew]  (1.5, 1.5) -- (2.5,1.75);
\node[bd] at (-1,0)  [label = below: \large{$1$}] {};
\node[bd] at (0,0) [label = below: \large{$2$}] {};
\node[bd] at (1,0)  [label = below: \large{$3$}] {};
\node[bd] at (0,2) [label = above: \large{$4$}] {};
\node[bd] at (0,1) {};
\end{tikzpicture} 
\ee
The set of $(p,q)$ charges is 
\be
W=
\begin{pmatrix}
 0 & 1 \\
 0 & 1 \\
 -2 & -1 \\
 2 & -1 \\
 \end{pmatrix}
\ee
The SNF results precisely in the $\mathbb{Z}_2$ 1-form symmetry. 
We see that there is only one set of $(p,q)$-charges with multiplicity bigger than 1: $(p,q)= (-1,0)$ and 
\be
m_{(-1,0)} = 2 \,.
\ee
Thus the reduced matrix is 
\be
W^{\text{red}_{(-1,0)}}=
\begin{pmatrix}
 -2 & -1 \\
 2 & -1 \\
 \end{pmatrix}
\ee
whose SNF is $\diag (4, 1)$, and thus confirms again the non-trivial extension $\mathcal{E}= \mathbb{Z}_4$, as expected \cite{Apruzzi:2021vcu}. 

\subsection{Equivalence to Intersection-Theoretic Approach to 2-Groups}\label{sec:yay!}

Here we show the equivalence of our method with the previous results that appeared in the literature about 2-groups \cite{Apruzzi:2021vcu}. In the context of the CY singularities we are considering in this paper as the main source of examples 
the key sequence \eqref{eq:boundary-bulk} reads 
\be
  \label{eq:boundary-bulk-CY}
  \ldots \to H_2(\mathcal X_\epsilon^6) \xrightarrow{i} H_2(\mathcal X_\epsilon^6, \mathbf{L}_{\mathcal X}^5  -
  \mathcal S) \to H_1(\mathbf{L}_{\mathcal X}^5 - \mathcal S) \to 0
\ee
where 
\be
  \widehat{\cE} = H_1(\mathbf{L}_{\mathcal X}^5 - \mathcal S) = \coker i \, .
\ee
In this case, we can give a second field theory interpretation of $\widehat{\cE}$ which recovers the previous results about 2-groups in the literature. 
By Lefschetz duality for triples \cite{Hatcher}, we have
$H_2(\mathcal X_\epsilon^6, \mathbf{L}_{\mathcal X}^5 - \mathcal S) = H^4(\mathcal X_\epsilon^6, \mathcal S)$, which by the
universal coefficient theorem, and the fact that
$H_3(\mathcal X_\epsilon^6, \mathcal S)=0$ since $\mathcal X$ is toric, is equal to
$\Hom(H_4(\mathcal X_\epsilon^6, \mathcal S), \bZ)$. The group $H_4(\mathcal X_\epsilon^6, \mathcal S)$ is
the group of compact divisors of $\mathcal X_\epsilon^6$, together with the
singular non-compact divisors, which give rise to relative 4-cycles in
the pair $(\mathcal X_\epsilon^6, \mathcal S)$. From this Lefschetz dual viewpoint, the
short exact sequence~\eqref{eq:boundary-bulk} becomes
\[
  \ldots H_2(\mathcal X_\epsilon^6) \xrightarrow{q} \Hom(H_4(\mathcal X_\epsilon^6, \mathcal S),
  \bZ) \to H_1(\mathbf{L}_{\mathcal X}^5 - \mathcal S) \to 0\, ,
\]
where the map $q$ is defined by the intersection pairing:
$(q(\Sigma_2))(D_4) = \Sigma_2\cdot D_4$. This intersection pairing
therefore measures the gauge and flavor charges of the dynamical
states. By exactness
\[
  \widehat{\cE} = H_1(\mathbf{L}_{\mathcal X}^5 - \mathcal S) = \coker q\, ,
\]
reproducing the prescription introduced for computing 2-groups in
\cite{Apruzzi:2021vcu}.  Indeed, the pairing $(q(\Sigma_2))(D_4)$ captures the intersection of  compact and non-compact divisors  with compact curves, i.e. 
\be
\mathcal{M} = \left(
\begin{array}{c} \\
\mathcal{M}_{4,2}^{G}  \\
\\
\hline\\
\mathcal{M}_{4,2}^{F}
\\
\, 
\\
\end{array}
\right) \,,
\ee
where the superscript specifies whether these are intersections with compact ($G$) or non-compact ($F$) divisors. The fact that we are computing the cokernel of $q$ then reproduces the prescription of \cite{Apruzzi:2021vcu}, namely
\be
\widehat{\mathcal{E}} = \mathbb{Z}^{r+f}/\mathcal{M}\mathbb{Z}^{r+f} \,,
\ee
thus showing that the formalism we developed in this paper correctly reproduces the previous results.

\section{Examples with 2-Group Symmetries}
\label{sec:Exes}

\subsection{Examples: 5d $SU(N)_k$ theories}

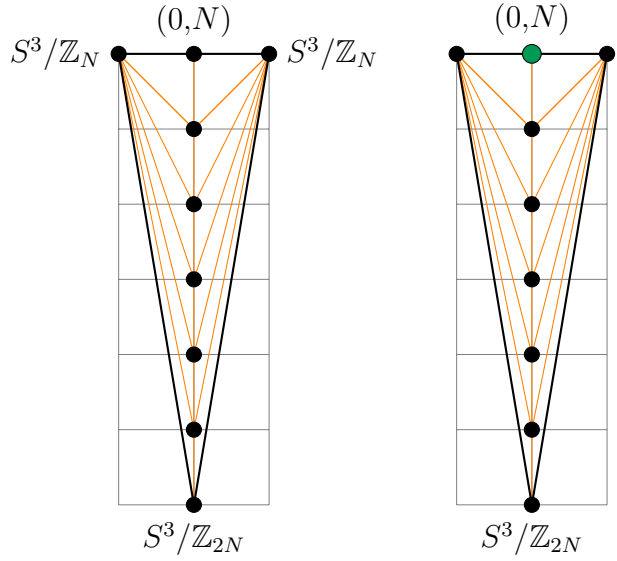
\begin{figure}\centering
\begin{tikzpicture}[x=1cm,y=1cm]
\draw[step=1cm,gray,very thin] (-1,0) grid (1,6);
\draw[lignet] (-1, 6) -- (0,5);
\draw[lignet] (-1, 6) -- (0,5);
\draw[lignet] (-1, 6) -- (0,4);
\draw[lignet] (-1, 6) -- (0,3);
\draw[lignet] (-1, 6) -- (0,2);
\draw[lignet] (-1, 6) -- (0,1);
\draw[lignet] (1, 6) -- (0,5);
\draw[lignet] (1, 6) -- (0,5);
\draw[lignet] (1, 6) -- (0,4);
\draw[lignet] (1, 6) -- (0,3);
\draw[lignet] (1, 6) -- (0,2);
\draw[lignet] (1, 6) -- (0,1);
\draw[lignet] (0,6) -- (0,0) ;
\draw[ligne] (-1,6)--(0,0)--(1,6)--(0,6)--(-1,6); 
\node[bd] at (0,0) [label=below: \large{$S^3/\mathbb{Z}_{2N}$}] {};; 
\node[bd] at (0,6) [label=above:\large{(0,$N$)}] {}; 
\node[bd] at (0,1) {}; 
\node[bd] at (0,2) {}; 
\node[bd] at (0,3) {};
\node[bd] at (0,4) {}; 
\node[bd] at (0,5) {}; 
\node[bd] at (1,6) [label=right:\large{$S^3/\mathbb{Z}_N$}]{}; 
\node[bd] at (-1,6) [label=left:\large{$S^3/\mathbb{Z}_N$}]{}; 
\end{tikzpicture} 
\qquad 
\begin{tikzpicture}[x=1cm,y=1cm]
\draw[step=1cm,gray,very thin] (-1,0) grid (1,6);
\draw[lignet] (-1, 6) -- (0,5);
\draw[lignet] (-1, 6) -- (0,5);
\draw[lignet] (-1, 6) -- (0,4);
\draw[lignet] (-1, 6) -- (0,3);
\draw[lignet] (-1, 6) -- (0,2);
\draw[lignet] (-1, 6) -- (0,1);
\draw[lignet] (1, 6) -- (0,5);
\draw[lignet] (1, 6) -- (0,5);
\draw[lignet] (1, 6) -- (0,4);
\draw[lignet] (1, 6) -- (0,3);
\draw[lignet] (1, 6) -- (0,2);
\draw[lignet] (1, 6) -- (0,1);
\draw[lignet] (0,6) -- (0,0) ;
\draw[ligne] (-1,6)--(0,0)--(1,6)--(0,6)--(-1,6); 
\node[bd] at (0,0) [label=below: \large{$S^3/\mathbb{Z}_{2N}$}] {};; 
\node[fd] at (0,6) [label=above:\large{(0,$N$)}] {}; 
\node[bd] at (0,1) {}; 
\node[bd] at (0,2) {}; 
\node[bd] at (0,3) {};
\node[bd] at (0,4) {}; 
\node[bd] at (0,5) {}; 
\node[bd] at (1,6) {}; 
\node[bd] at (-1,6) {}; 
\end{tikzpicture} 
\caption{The toric diagram from $SU(N)_N$ shown here on the left for $SU(6)_6$, which has  $\mathbb{Z}_6$ 1-form symmetry. The orange lines indicate the triangles which determine the order of the $S^3/\mathbb{Z}_k$ quotient. 
The right figure shows the toric diagram after we excise the $A_1$ singularity (shown in green), which is associate to the $\mathcal{F}=SO(3)$ flavor symmetry group. The only remaining lens space is the $S^3/\mathbb{Z}_{2N}$. 
For even $N$ this forms a non-trivial extension with the 1-form symmetry $\mathbb{Z}_N$ and thereby a 2-group. 
 \label{fig:ToricSUpq}}
\end{figure}

For pure gauge theories in 5d with gauge group $SU(N)$ and CS-level $k$ it is known \cite{Apruzzi:2021vcu} that the theories with 2-groups are 
\be
SU(2n)_{2n} :\qquad \Gamma^{(1)}= \mathbb{Z}_{2n} \,, \qquad \mathcal{F} = SO(3) \,.
\ee
The theories with other CS-levels or $SU(2n+1)$ have trivial 2-groups. 
The toric diagram for one of these is e.g. in figure \ref{fig:ToricSUpq}. From the left hand figure we infer that the one-form symmetry is indeed $\mathbb{Z}_{\text{gcd}(N, N, 2N)} = \mathbb{Z}_{N}$, whereas from the right hand side, after we excise the flavor symmetry vertex, the only remaining lens space singularity at the boundary is $S^3/\mathbb{Z}_{2N}$. For $N= 2n$ these fit into the non-split short exact sequence
\be
1\ \rightarrow \ \mathbb{Z}_{2n} \ \rightarrow \ (\mathcal{E}= \mathbb{Z}_{4n}) \ \rightarrow \ \mathbb{Z}_2 \rightarrow 1\,.
\ee

We can of course also apply the approach using the dual webs in section \ref{sec:Webs}. 
First of all it is obvious that for $k\not N$ there is no 2-group (since there is no non-abelian flavor symmetry, and thus no multiplicities in the $(p,q)$-charge matrix $W$). For $k=N$, 
\be
W= \begin{pmatrix}
0&1 \\ 
0&1 \\
1& -N \\
1 & N  
\end{pmatrix}
\ee
The Smith normal form of $W$ is diag$(1,N)$, and thereby $\Gamma^{(1)}= \mathbb{Z}_N$. 
To compute the extension group $\mathcal{E}$, note that the entry $(-1,0)$ has multiplicities $m_{(-1,0)}=2$ and so 
\be
W^{\text{red}}_{(-1,0)} = 
\begin{pmatrix}
1& -N \\
1 & N  
\end{pmatrix}
\ee
whose Smith normal form is 
\be
\text{SNF}  (W^{\text{red}}_{(-1,0)}) = \text{diag} \left(1,  2^{(1+(-1)^N)/2} N  \right) \,.
\ee
So that $\mathcal{E} = \mathbb{Z}_{2N}$ for $N$ even and $\mathbb{Z}_{N}$ for $N$ odd.

\subsection{Quivers}\label{sec:quivers}

There are numerous quiver theories which have 2-groups. We select out the following class (which are closely related to $SU(N)_k + 1\bm{AS}$, as we will discuss later). The new feature in this class of examples is that the flavor symmetry has multiple components. 

\begin{figure}\centering
\begin{tikzpicture}[x=1cm,y=1cm]
\draw[step=1cm,gray,very thin] (0,-2) grid (3,2);
%\draw[lignet] (1,-1) -- (3,-1);
%\draw[lignet] (1,-1) -- (1,1);
%\draw[lignet] (3,-1) -- (1,1);
 \draw[ligne] (0,0)-- (1,-1)--(2,-2)--(3,-1)--(2,2)--(1,1)--(0,0); 
\node[bd] at (0,0) [label=left:\large{$S^3/\mathbb{Z}_2$}] {}; 
\node[bd] at (1,-1) [label=left:\large{$2$}]{};
\node[bd] at (2,-2)[label=below:\large{$S^3/\mathbb{Z}_2$}] {};
\node[bd] at (3,-1)[label=right:\large{$S^3/\mathbb{Z}_4$}] {};
\node[bd] at (2,2) [label=above:\large{$S^3/\mathbb{Z}_4$}]{};
\node[bd] at (1,1)  [label=left:\large{$6$}]{};
\node[bd] at (1,0) {};
\node[bd] at (2,0) {};  
\node[bd] at (2,1) {};  
\node[bd] at (2,2) {};  
\node[bd] at (2,-1) {};  
\end{tikzpicture}
\caption{The toric diagram for $SU(2)_0 - SU(4)_0$ quiver. 
 \label{fig:SU2SU4}}
\end{figure}
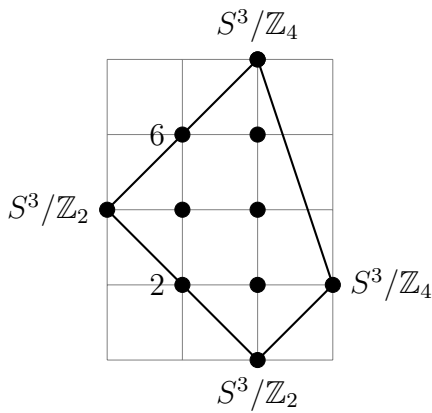

Let us first discuss the simplest example. Consider the toric diagram drawn in figure \ref{fig:SU2SU4}. This theory has a description in terms of the strongly-coupled limit of an $SU(2)_0 - SU(4)_0$ quiver gauge theory. 
Figure \ref{fig:SU2SU4} shows already the asymptotic lens spaces, which imply the 1-form symmetry is $\Gamma^{(1)}=\mathbb{Z}_2$.

To compute the 2-group structure, note that there are two edges with vertices along them: $v_2$ and $v_6$ respectively. We now excise these in turn. Excision of the vertex $v_2$ (and thereby the edge $e=1$), results in 
$\mathcal{E}_{e=1} = \mathbb{Z}_4$, so this participates in a 2-group, whereas excision of $v_6$ results in  $\mathcal{E}_{e=4}= \mathbb{Z}_2$, and this flavor symmetry will not participate in the 2-group. 
The flavor symmetry group of this theory is 
\be
\mathcal{F}= SO(3) \times SU(2)
\ee
and the first factor takes part in a non-trivial 2-group structure.

Using the brane-web approach we find the following brane-charges
\be
W= \begin{pmatrix}
1 & 1 \\
 1 & 1 \\
 -1 & 1 \\
 -3 & -1 \\
 1 & -1 \\
 1 & -1 \\
\end{pmatrix} \,,
\ee
from where we find the 1-form symmetry $\Gamma^{(1)}= \mathbb{Z}_2$. 
The flavor symmetry algebra can be read off from the toric diagram to be 
\be
\mathfrak{f} = \mathfrak{su}(2)^{(1)} \oplus \mathfrak{su}(2)^{(2)} \,,
\ee
where the first is associated to the multiplicity $m_{(-1,1)} = 2$ and the second to $m_{(1,1)} =2$. 
From the reduced W-matrices we obtain 
\be
\SNF\left(W^{\text{red}}_{(-1, 1)}\right) = \diag (1, 4) \,,\qquad 
\SNF\left(W^{\text{red}}_{(1, 1)}\right) = \diag (1, 2) \,,\qquad 
\ee
so that there is a non-trivial extension, where however only the first $\mathfrak{su}(2)$ factor participates in.

\subsection{Non-Lagrangian Theories with 2-Groups}

\begin{figure}\centering
\begin{tikzpicture}[x=1cm,y=1cm]
\draw[step=1cm,gray,very thin] (0,0) grid (4,3);
\draw[lignet] (0,2) -- (4,0);
\draw[lignet] (0,2) -- (1,1);
\draw[lignet] (1,1) -- (4,0);
\draw[lignet] (0,1) -- (2,1);
\draw[lignet] (0,3) -- (1,2);
\draw[lignet] (0,3) -- (2,1);
\draw[lignet] (0,2) -- (1,2);
\draw[lignet] (1,2) -- (4,0);
\draw[ligne] (4,0)--(0,3)--(0,2)--(0,1)--(4,0); 
\node[bd] at (4,0) [label=right:\large{$S^3/\mathbb{Z}_8$}] {}; 
\node[bd] at (0,3) [label=left:\large{$S^3/\mathbb{Z}_4$}]{};
\node[bd] at (0,2) {};
\node[bd] at (0,1) [label=left:\large{$S^3/\mathbb{Z}_4$}]{};
\node[bd] at (1,1) {};
\node[bd] at (1,2) {};
\node[bd] at (2,1) {};
\node[wd] at (0,0) [label=below:\large{(0,0)}]  {}; 
\end{tikzpicture}
\qquad 
\begin{tikzpicture}[x=1cm,y=1cm]
\draw[step=1cm,gray,very thin] (0,0) grid (4,3);
\draw[lignet] (0,2) -- (4,0);
\draw[lignet] (0,2) -- (1,1);
\draw[lignet] (1,1) -- (4,0);
\draw[lignet] (0,1) -- (2,1);
\draw[lignet] (0,3) -- (1,2);
\draw[lignet] (0,3) -- (2,1);
\draw[lignet] (0,2) -- (1,2);
\draw[lignet] (1,2) -- (4,0);
\draw[ligne] (4,0)--(0,3)--(0,2)--(0,1)--(4,0); 
\node[bd] at (4,0) [label=right:\large{$S^3/\mathbb{Z}_8$}] {}; 
\node[bd] at (0,3) {};
\node[fd] at (0,2) {};
\node[bd] at (0,1) {};
\node[bd] at (1,1) {};
\node[bd] at (1,2) {};
\node[bd] at (2,1) {};
\node[wd] at (0,0) [label=below:\large{(0,0)}]  {}; 
\end{tikzpicture}
\caption{The toric diagram from $B_4^{(2)}$. 
 \label{fig:B42}}
\end{figure}
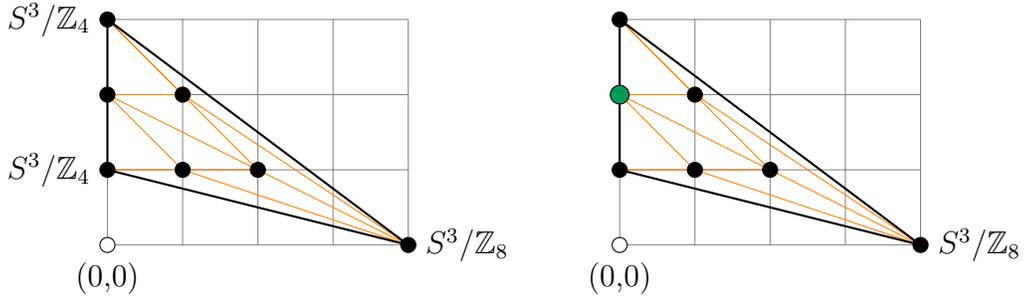

We now construct non-Lagrangian theories with 2-group symmetries. 
Consider the toric diagrams defined by \cite{Eckhard:2020jyr, Morrison:2020ool}: 
\be
B_N^{(2)} :\qquad  ( (N,0, 1), (0, N-1-k, 1)), \, k= 0, \cdots, N-2  \,,
\ee
which have one-form symmetry $\Gamma^{(1)}= \mathbb{Z}_{N}$ and flavor symmetry group $\mathcal{F}= SU(N-2)/\mathbb{Z}_{N-2}$.
In figure  \ref{fig:B42} we show the example for $B_4^{(2)}$. In this case we find that indeed 
\be
B_4^{(2)}:\qquad \Gamma^{(1)} = \mathbb{Z}_{\text{gcd}(4, 4, 8)} = \mathbb{Z}_{4} \,,
\ee
and the 2-group symmetry arises after excising the $SO(3)$ flavor node. 
In general we find: 
\be
B_{N}^{(2)}:\qquad \Gamma^{(1)} = \mathbb{Z}_{\text{gcd}(N, N, (N-2) N )} = \mathbb{Z}_{N} \,.
\ee
Excising the flavor vertices associated to $\mathcal{F}$ we find 
\be
\mathcal{E} = \mathbb{Z}_{N (N-2)}\,,
\ee
and thus there is a non-trivial extension only for $N=2n$.

%\begin{table}
%$$
%\begin{array}{|c|c|c|}\hline
%\text{Theory} & \Gamma^{(1)} & \text{Toric Fan} \cr \hline\hline
%B_N & \mathbb{Z}_{N (N-3)+ 3}&  (N-1, 0,1), (1, N-1, 1), (0,1,1)\cr 
%B_N^{(1)} & \mathbb{Z}_{N-1} &   ((N-1,0, 1), (N-1, 1, 1) (0, N-1-k, 1)), \, k= 0, \cdots, N-2  \cr 
%B_N^{(2)} & \mathbb{Z}_{N} &  ( (N,0, 1), (0, N-1-k, 1)), \, k= 0, \cdots, N-2  \cr 
%B_N^{(3)} & \mathbb{Z}_{N-1} &  ( (0,1, 1),   (N-k, k, 1)), \, k= 0, \cdots, N-1   \cr \hline
%\end{array}$$
%\caption{Properties of the $B_N$ and $B_N^{(i)}$ non-Lagrangian toric models. \label{tab:BN}}
%\end{table}

%
%\newpage
%%%%%%%%%%%%%%%%%%%%%%%
%\begin{figure}\centering
%\begin{tabular}{cc}
%\begin{tikzpicture}[x=1cm,y=1cm]
%\draw[step=1cm,gray,very thin] (0,0) grid (3,3);
%\node[] at (1.5,3.5) {$B_4$};
%\draw[ligne] (3,0)--(1,3)--(0,1)-- (3,0); 
%\node[bd] at (3,0) {}; 
%\node[bd] at (1,3) {}; 
%\node[bd] at (0,1) {};
%\node[bd] at (1,1) {};
%\node[bd] at (1,2) {};
%\node[bd] at (2,1) {};
%\node[wd] at (0,0) [label=below:\large{(0,0)}]  {}; 
%\end{tikzpicture} 
%& \qquad
% \cr 
%\begin{tikzpicture}[x=1cm,y=1cm]
%\draw[step=1cm,gray,very thin] (0,0) grid (3,3);
%\node[] at (1.5,3.5) {$B_4^{(1)}$};
%\draw[ligne] (3,0)--(3,1)--(0,3)--(0,2)--(0,1)--(3,0); 
%\node[bd] at (3,0) {}; 
%\node[bd] at (3,1) {}; 
%\node[bd] at (0,3) {};
%\node[bd] at (0,2) {};
%\node[bd] at (0,1) {};
%\node[bd] at (1,1) {};
%\node[bd] at (1,2) {};
%\node[bd] at (2,1) {};
%\node[wd] at (0,0) [label=below:\large{(0,0)}]  {}; 
%\end{tikzpicture} 
%&\qquad
%\begin{tikzpicture}[x=1cm,y=1cm]
%\draw[step=1cm,gray,very thin] (0,0) grid (4,3);
%\node[] at (1.5,3.5) {$B_4^{(3)}$};
%\draw[ligne] (4,0)--(3,1)--(2,2)--(1,3)--(0,1) -- (4,0); 
%\node[bd] at (4,0) {}; 
%\node[bd] at (3,1) {};
%\node[bd] at (2,2) {};
%\node[bd] at (1,3) {};
%\node[bd] at (0,1) {};
%\node[bd] at (1,1) {};
%\node[bd] at (1,2) {};
%\node[bd] at (2,1) {};
%\node[wd] at (0,0) [label=below:\large{(0,0)}]  {}; 
%\end{tikzpicture} 
%\end{tabular}
%\caption{$B_N$ and $B_N^{(i)}$ non-Lagrangian toric diagrams for $N=4$. \label{fig:BN}}
%\end{figure}

\subsection{Generalized Toric Geometry}
\label{sec:beyondtoric}

Although the precise geometric meaning of generalized toric diagrams (GTP) \cite{Benini:2009gi, vanBeest:2020kou, VanBeest:2020kxw}
still remains to be understood, we can nevertheless apply our approach to the dual brane-webs. 
Consider a GTP, and let $P$ be the toric polygon where all white-dots are replaced by black dots. 
Let $W$ again be the set of $(p,q)$ brane-charges associated to $P$ (i.e. not the GTP),
then the 1-form symmetry is computed from the SNF of $P$ \cite{Bhardwaj:2020phs}. 
Let $W$ be the set brane-charges corresponding to $P$.

Conjecturally we find the following rule: To compute the 1-form symmetry, compute as before in the toric case $\SNF(W)$. 
For the 2-group, we again excise the $(p,q)$-charges which have non-trivial multiplicities in the original GTP (not the one where all white dots have been replaced with black).
Consider first the theory $SU(4)_0 + 1\bm{AS}$, which is closely related to the toric quivers we discussed earlier. 
It has $\Gamma^{(1)}= \mathbb{Z}_2$ and GTP 
\be
\begin{tikzpicture}[x=1cm,y=1cm]
\draw[step=1cm,gray,very thin] (0,-2) grid (3,2);
\draw[ligne] (0,0)-- (1,-1)--(2,-2)--(3,-1)--(2,2)--(1,1)--(0,0); 
\node[bd] at (0,0) {}; 
\node[wd] at (1,-1) {};
\node[bd] at (2,-2) {};
\node[bd] at (3,-1) {};
\node[bd] at (2,2) {};
\node[bd] at (1,1) {};
\end{tikzpicture}
\ee
The associated brane-charges $W^{\text{GTP}}$ for the GTP, and $W$-matrix for the ``filled'' web are 
\be
W^{\text{GTP}}= 
\begin{pmatrix}
 2 & 2 \\
 -1 & 1 \\
 -3 & -1 \\
 1 & -1 \\
 1 & -1 \\
\end{pmatrix}
\,,\qquad W= \begin{pmatrix}
1 & 1 \\
 1 & 1 \\
 -1 & 1 \\
 -3 & -1 \\
 1 & -1 \\
 1 & -1 \\
\end{pmatrix}
\ee
The only multiplicity that does not correspond to a white-dot is $m_{(1,1)}=2$ and 
\be
\SNF (W_{(1,1)}^{\text{red}}) = \diag (1,2) \,,
\ee
so there is no 2-group, consistent with \cite{Apruzzi:2021vcu}. 

On the other hand $SU(4)_2 + 1\bm{AS}$ has $\Gamma^{(1)}= \mathbb{Z}_2$ and GTP 
\be
\begin{tikzpicture}[x=1cm,y=1cm]
\draw[step=1cm,gray,very thin] (0,-2) grid (3,2);
\draw[ligne] (0,0)-- (1,-1)--(2,-2)--(3,1)--(2,2)--(1,1)--(0,0); 
\node[bd] at (0,0) {}; 
\node[wd] at (1,-1) {};
\node[bd] at (2,-2) {};
\node[bd] at (3,1) {};
\node[bd] at (2,2) {};
\node[bd] at (1,1) {};
\end{tikzpicture}
\ee
with 
\be
W^{\text{GTP}}= 
\begin{pmatrix}
2 & 2 \\
 -3 & 1 \\
 -1 & -1 \\
 1 & -1 \\
 1 & -1 \\
 \end{pmatrix}
 \,,\qquad W= \begin{pmatrix}
 1 & 1 \\
 1 & 1 \\
 -3 & 1 \\
 -1 & -1 \\
 1 & -1 \\
 1 & -1 \\
  \end{pmatrix}
\ee
Again the only non-white dot multiplicity is $m_{(1,1)}=2$ but now 
\be
\SNF (W_{(1,1)}^{\text{red}}) = \diag (1,4) \,,
\ee
indicating a non-trivial extension, again consistent with \cite{Apruzzi:2021vcu}. 
Clearly applying this to further toric and generalized toric models is straight forward, and only requires the flavor symmetry to be manifest in the geometric description.

\subsection*{Acknowledgements}
 We thank Fabio Apruzzi, Lakshya Bhardwaj, Antoine Bourget, Jonathan Heckman, Saghar S. Hosseini, Max Hubner, Dave Morrison, Mark Powell and Yi-Nan Wang for discussions. SSN is supported in part by the European Union’s Horizon 2020 Framework: ERC grant 682608 and in part by the “Simons Collaboration on Special Holonomy in Geometry, Analysis and Physics” grant number 724073. MDZ acknowledges support from the European Research Council (ERC) under the European Union’s Horizon 2020 research and innovation programme (grant agreement No. 851931). MDZ and IGE are supported in part by the ``Simons Collaboration on Global Categorical Symmetries.'' IGE is further supported in part by STFC grant ST/T000708/1.

%%%%%%%%%%%%%%%%%%%%%%%%%%%%%%%%%%%%%%%%%%%%%%

\bibliographystyle{JHEP}

\bibliography{refs}

\end{document}